\documentclass[twocolumn,aps,pra,superscriptaddress,10pt]{revtex4-1}
\usepackage{graphicx}
\usepackage{amsmath,amsfonts,amsthm,bm}
\usepackage{amssymb}
\usepackage{color} 
\usepackage{braket}
\usepackage{bm}
\usepackage{dcolumn}
\usepackage{subfigure}
\usepackage[colorlinks=true,linkcolor=blue,citecolor=blue,urlcolor=blue]{hyperref}
\usepackage{soul}
\usepackage{changes}
\usepackage{mathtools}
\usepackage{mathrsfs}

\usepackage{adjustbox}
\usepackage{multirow}

\makeatother

\begin{document}

\title{Universal energy-dependent pseudopotential for the two-body problem of
confined ultracold atoms}

\author{Da-Wu Xiao}

\affiliation{Beijing Computational Science Research Center, Beijing, 100193, China}

\author{Ren Zhang}

\affiliation{School of Physics, Xi'an Jiaotong University, Xi'an, 710049, China}

\author{Peng Zhang}
\email{pengzhang@ruc.edu.cn}

\affiliation{Department of Physics, Renmin University of China, Beijing, 100872, China}

\affiliation{Beijing Computational Science Research Center, Beijing, 100193, China}

\date{\today}
\begin{abstract}
The two-body scattering amplitude and energy spectrum of confined
ultracold atoms are of fundamental importance for both theoretical
and experimental studies of ultracold atom physics. For many systems, one can
efficiently calculate these quantities via the
zero-range Huang-Yang pseudopotential (HYP),
in which the interatomic interaction is characterized by the
scattering length $a$. Furthermore, when the scattering length is \textit{dependent
on the kinetic energy $\varepsilon_{\rm r}$ of two-atom relative motion}, i.e., $a=a(\varepsilon_{\rm r})$, the results are
applicable for a broad energy region.
However, when the free Hamiltonian of atomic internal state (\textit{e.g.}, the Zeeman Hamiltonian) does not commute with the inter-atomic interaction, or the center-of-mass (c.m.) motion is coupled to the relative motion, the generalization of this technique is still lacking.
In this work we solve this problem and construct a reasonable energy-dependent multi-channel HYP, which is characterized by a ``scattering length operator" ${\hat a}_{\rm eff}$, for the above complicated cases. Here ${\hat a}_{\rm eff}$ is an operator for atomic internal states and c.m. motion, and depends on both the total two-atom energy and the external field as well as the trapping parameter.
The effects from the internal-state or
c.m.-relative motion coupling can be self-consistently taken into account by ${\hat a}_{\rm eff}$. We further show a method based on the quantum defect theory, with which
 ${\hat a}_{\rm eff}$ can be analytically derived for
systems with van der Waals inter-atomic interaction.
To demonstrate our method, we
calculate the spectrum of two
ultracold fermionic 
alkaline-earth-like
atoms (in electronic $^{1}{\rm S}_{0}$ ($|g\rangle$) and $^{3}{\rm P}_{0}$ ($|e\rangle$) states, respectively) confined in an optical lattice.
By comparing our results with the recent experimental measurements 
for two $^{173}$Yb atoms and two $^{171}$Yb atoms, we calibrate the scattering lengths
$a_{\pm}$ with respect to anti-symmetric and symmetric nuclear-spin states
to be $a_{+}=2012(19)a_{0}$ and $a_{-}=193(4)a_{0}$ for $^{173}$Yb, and
$a_{+}=232(3)a_{0}$ and $a_{-}=372(1)a_{0}$ for $^{171}$Yb.
\end{abstract}
\maketitle

\section{introduction}
\label{sec:sec1}

Two-body physics of ultracold atoms
in various confinements plays a very basic and important role
in the studies of ultracold gases \cite{RMP2010,FRRMP}. For instance, the effective inter-atomic interaction of
ultracold atoms in quasi low- (or mixed-) dimensional confinements is characterized by the two-body
scattering amplitude. As a result, one can control this effective
pairwise interaction by tuning the scattering amplitude through the confinement
parameter \cite{OshaniiCIR,TanShinaCIR,ZhangCIR,ZhangPRA2016,ZhangPRA2017,RieggerPRL2018,ZhangPRA2020,PengSGPRA,PeanoNJP2005,CIRPeng2011,SalaPRL2012,SalaPRL2013,MelezhikPRA2011,GiannakeasPRA2012,BenjaminPRA2014,Wang2016,GassabPRA2015}. In addition, using the energy spectrum of two ultracold
atoms in a three-dimensional (3D) confinement, one can not only qualitatively
obtain a primary understanding for the interacting physics, but
also quantitatively calculate some important physical parameters
for the many-body physics, such as the second virial coefficient \cite{LIU201337,PSGPRA}. Moreover,
the systems of two ultracold atoms in 3D confinements have already
been realized in many recent experiments, and the two-body energy spectrum and dynamics
can be directly observed \cite{PRL2015,PRX2019,Bettermann,PRL2014,PRA2019,Scazza_2014,Benjamin}. 
These observations also call for a deep understanding of two-body physics.

To calculate the two-atom scattering amplitude or energy spectrum,
one needs to solve the Schr\"odinger equation for two interacting
ultracold atoms in confinements. For ultracold atoms, one can
ignore the short-range details of the bare inter-atomic interaction $U_{{\rm bare}}(r)$, with
$r=|{\bf r}|$ and ${\bf r}$ being the relative position of these two atoms,
 and approximate this interaction
with the zero-range effective potential. As such, the calculations for the two-atom scattering amplitude or energy spectrum
can be significantly simplified.
In the zero-energy limit, a prevailing effective interaction potential is the Huang-Yang pseudopotential (HYP)
\begin{eqnarray}
U_{{\rm HY}}(a)\equiv\frac{2\pi\hbar^{2}a}{\mu}\delta({\bf r})\frac{\partial}{\partial r}(r\cdot),
\end{eqnarray}
where $\mu$ is the reduced mass,
and the energy-independent parameter $a$ is the $s$-wave
scattering length, which is determined by the zero-energy scattering amplitude of these two atoms in 3D free space.

In many cases the finite-energy effect of the  
scattering is required to be taken into account. One simple approach
to achieve this goal is to use the ``energy-dependent HYP\char`\"{}
$U_{{\rm HY}}[a(\varepsilon_{{\rm r}})]$, where $a$ is replaced
by an \textit{``energy-dependent scattering length\char`\"{}} $a(\varepsilon_{{\rm r}})$,
which is defined as
\begin{eqnarray}
a(\varepsilon_{{\rm r}})=-\frac{\tan\delta_{s}(\varepsilon_{{\rm r}})}{\sqrt{2\mu\varepsilon_{{\rm r}}/\hbar^2}},\label{ae}
\end{eqnarray}
with $\delta_{s}(\varepsilon_{{\rm r}})$ being the $s$-wave phase
shift for the scattering in 3D free space, with respect to finite
scattering energy $\varepsilon_{{\rm r}}$ \cite{footnote,footnote3}. For the confined ultracold
atoms discussed in this work, the energy $\varepsilon_{{\rm r}}$
is 
the kinetic energy of the two-atom relative motion in the
short-range region $d_{\rm int}\lesssim r\ll d_{\rm trap}$, with $d_{\rm int}$ and $d_{\rm trap}$ being the characteristic lengths
of the range of $U_{{\rm bare}}(r)$ and the confinement, respectively, and the condition $d_{\rm int}\ll d_{\rm trap}$
is satisfied in
almost of all the current experiments.
Therefore, when the atoms are single-component and
the relative motion is decoupled from the center-of-mass
(c.m.) motion, $\varepsilon_{{\rm r}}$ can be simply re-expressed as $\varepsilon_{{\rm r}}=E-V^{\rm (c)}({\bf r}={\bf 0})$, with $V^{\rm (c)}({\bf r})$ being the confinement-contributed potential-energy
term in the two-atom relative Hamiltonian, and $E$ is the total energy
of the two-atom relative motion. Thus, one
can calculate the two-body scattering amplitude or energy spectrum
by self-consistently solving the stationary Schr\"odinger equation
\begin{eqnarray}
 & & \left\{ -\frac{\hbar^{2}}{2\mu}\nabla_{{\bf r}}^{2}+U_{{\rm HY}}\Big[a\Big(E-V^{{\rm (c)}}({\bf r}={\bf 0})\Big)\Big]+V^{{\rm (c)}}({\bf r})\right\} \psi({\bf r})\nonumber \\
\nonumber \\
 & &= E\psi({\bf r}).
\end{eqnarray}

On the other hand, in experiments, there are also various relatively complicated systems with {at
least one of the following two situations:}
\begin{itemize}
\item[(A)] The atomic relative and c.m. motions are coupled.

\item[(B)] The atoms are multi-component, and the inter-atomic interaction does not commute with the
free Hamiltonian of the internal state ({\it e.g.}, the Zeeman Hamiltonian).
\end{itemize}
The examples of systems with situation (A) include ultracold atoms in anharmonic confinement, and two
heteronuclear
atoms in species-dependent confinement.
The examples of systems with the situation
(B) are two confined homonuclear fermionic
alkaline-earth (like) atoms in $^{1}{\rm S}_{0}$ and $^{3}{\rm P}_{0}$
states, which are subjected to a Zeeman magnetic field. In the latter system, each atom can be in several different nuclear-spin states, and the
 $s$-wave inter-atomic interaction is diagonal in the basis of symmetric and anti-symmetric nuclear-spin states, and is not commutative with the Zeeman Hamiltonian~ \cite{YeJunScience,Scazza_2014,AnaMariaNP}.

In the presence of situation (A) or (B), one cannot directly use
the above simple approach of HYP with energy-dependent scattering lengths. That is because:
when these situations arise the relative motion of the two atoms would
be entangled with the atomic internal states or the c.m. motion. As
a result, the kinetic energy $\varepsilon_{{\rm r}}$ of the relative
motion does not take a definite value, even in the short-range region.
Therefore, the function $a(\varepsilon_{{\rm r}})$ does not have a specific argument value, and thus this function cannot
be directly applied. To solve
the two-body problems with the above two situations, one has to either
completely ignore the finite-energy effect of the 3D scattering \cite{PeanoNJP2005,CIRPeng2011}, or solve
the Schr\"odinger equation with both the confinement potential and a
more complicated inter-atomic interaction model,
such as a finite-range model \cite{MelezhikPRA2011,SalaPRL2012,SalaPRL2013,SalaPRA2016} or a zero-range model with auxiliary closed channels (auxiliary molecule channels). Notice that the latter one cannot be used in the presence of the situation(B), as shown below.

In this work we solve this problem by constructing an energy-dependent
HYP
\begin{eqnarray}
{\hat U}_{\rm eff}(E)={\hat a}_{\rm eff}(E)\frac{2\pi\hbar^{2}}{\mu}\delta({\bf r})\frac{\partial}{\partial r}(r\cdot),\label{ue2}
\end{eqnarray}
 for systems with situations (A) or (B) or both,
 which is characterized by a ``scattering length operator"
 ${\hat a}_{\rm eff}$.
 Here $E$ is the total energy of this two-body system, and ${\hat a}_{\rm eff}$ is an operator
 of the Hilbert space of two-atom
internal state or c.m. motion, which depends on not only the energy $E$ but also the external field and the confinement potential.
The effects induced by the situations (A) and (B) can be self-consistently encapsulated by ${\hat a}_{\rm eff}$.
For most systems ${\hat a}_{\rm eff}$ cannot be obtained with simple transformations on the single-channel energy-dependent scattering length $a(\varepsilon_{\rm r})$. We show the approach to derive ${\hat a}_{\rm eff}$ for general cases, and further develop a multi-channel quantum-defect theory (MQDT) \cite{Gao2007,Zhang2017,Gao2005,Gao1998QDT} with which one can analytically calculate all the matrix-elements of ${\hat a}_{\rm eff}$, for systems where $U_{\rm bare}(r)$ can be approximated as an internal-state independent van der Waals potential for $r>b$, with $b$ being a particular range.

The HYP ${\hat { U}_{\rm eff}(E)}$ we developed can be used for the
calculations of two-body scattering amplitude or energy spectrum. The
calculations (including the ones to derive ${\hat a}_{\rm eff}$) are much simpler in comparison with the ones with a
finite-range interaction model or auxiliary closed channels.

As a demonstration, we calculate the energy spectrum
of two homonuclear fermionic alkaline-earth (like) atoms confined in a site of an optical lattice, which
are in electronic $^{1}{\rm S}_{0}$ and $^{3}{\rm P}_{0}$
states, respectively, in the presence of a Zeeman field.
As mentioned above, the $s$-wave interaction between these two atoms is
diagonal in the basis of anti-symmetric and symmetric nuclear-spin states, and is characterized by the zero-energy scattering lengths $a_{+}$ and $a_{-}$ of the corresponding potential curves \cite{YeJunScience,Scazza_2014,AnaMariaNP}. As a result, there are nuclear-spin exchange interactions
between these two atoms, with the intensity being proportional
to $(a_{-}-a_{+})$ in the zero-range limit \cite{AnaMariaNP,Scazza_2014,PRA2019,PRL2014,Benjamin}. Thus, the mixture of ultracold atoms in $^{1}{\rm S}_{0}$ and $^{3}{\rm P}_{0}$states is a promising
candidate for the quantum simulation of many-body physics induced
by spin-exchange interaction (\textit{e.g.}, the Kondo physics), and
has attracted much attention \cite{AnaMariaNP,ZhangPRA2016,Zhang_2020,ZhangPRA2017,ZhangPRA2020,PRA2019,KondoPRL,KondoPRL2016,KondoPRL2015,KondoPRA2010}. Furthermore, the precise values of $a_{\pm}$ are required as basic parameters
for the study of this quantum simulation.
For $^{173}{\rm Yb}$ and $^{171}{\rm Yb}$ atoms these values have
been derived by several groups via comparing the experimentally-measured
the energy spectrum of two atoms confined in an optical lattice site with corresponding
theoretical calculations \cite{PRL2015,PRX2019,Bettermann}. 
These experiments were done under a finite
Zeeman magnetic field. However, in the previous calculations the
Zeeman energies were ignored in the short-range region, which is actually
non-negligible. In this work, we calibrate the values of $a_{\pm}$
by fitting the energy spectrum calculated via our approach,
where the Zeeman coupling in all the spatial space is included, with
the experimental measurements. We obtain the calibrated value $a_{+}=2012(19)a_{0}$
and $a_{-}=193(4)a_{0}$ for $^{173}$Yb, and $a_{+}=232(3)a_{0}$ and $a_{-}=372(1)a_{0}$
for $^{171}$Yb, which are at most 12\% different from the ones given
by previous works (Table \ref{tab:tabI}).

{\renewcommand{\arraystretch}{1.5}
\begin{table}[ptb]
 \caption{The fitting scattering lengths of ultracold $^{173}\rm{Yb}$ and $^{171}\rm{Yb}$ atoms, where $a_0$ is the Bohr radius.
 }
 \begin{ruledtabular}
{\centering
 \begin{tabular}{cccccc}

   \multirow{3}{*}{$^{173}\rm{Yb}$} & \multicolumn{1}{c}{Ref.} & \multicolumn{1}{c}{} & \multicolumn{1}{c}{\cite{PRL2015}} & \multicolumn{1}{c}{\cite{PRX2019}} & \multicolumn{1}{c}{This work}\\\cline{2-6}
         & \multicolumn{1}{c}{$a_+/a_0$} & \multicolumn{1}{c}{} & \multicolumn{1}{c}{1878} & \multicolumn{1}{c}{1894(18)} & \multicolumn{1}{c}{2012(19)}\\\cline{2-6}
         & \multicolumn{1}{c}{$a_-/a_0$} & \multicolumn{1}{c}{} & \multicolumn{1}{c}{219.7} & \multicolumn{1}{c}{/} & \multicolumn{1}{c}{193(4)}\\\cline{2-6}
   \hline \hline
   \multirow{3}{*}{$^{171}\rm{Yb}$} & \multicolumn{1}{c}{Ref.} & \multicolumn{1}{c}{} & \multicolumn{1}{c}{\cite{Bettermann}} & \multicolumn{1}{c}{} & \multicolumn{1}{c}{This work}\\\cline{2-6}
         & \multicolumn{1}{c}{$a_+/a_0$} & \multicolumn{1}{c}{} & \multicolumn{1}{c}{240(4)} & \multicolumn{1}{c}{} & \multicolumn{1}{c}{232(3)}\\\cline{2-6}
         & \multicolumn{1}{c}{$a_-/a_0$} & \multicolumn{1}{c}{} & \multicolumn{1}{c}{389(4)} & \multicolumn{1}{c}{} & \multicolumn{1}{c}{372(1)}\\

\end{tabular}
}
 \end{ruledtabular}
 \label{tab:tabI}
\end{table}
}

The remainder of this paper is organized as follows. In Sec.~\ref{sec:sec2} we
show our approach for the construction of the scattering length operator ${\hat a}_{\rm eff}$
 for general cases.
In Sec.~\ref{sec:sec3} we demonstrate this approach with the calculation of the energy spectrum of
two confined alkaline-earth (like) atoms. We illustrate our results for
two $^{173}$Yb atoms and two $^{171}$Yb atoms
and calibrate the
scattering lengths in Sec.~\ref{sec:sec4}. A summary of our work and the outlook
of our method are given in Sec.~\ref{sec:sec5}. Some details of our calculations are given
in the appendixes.

\section{General Approach}
\label{sec:sec2}

\subsection{System and basic idea}
\label{sec:sec2A}

\begin{figure}[b]
\includegraphics[scale=0.45]{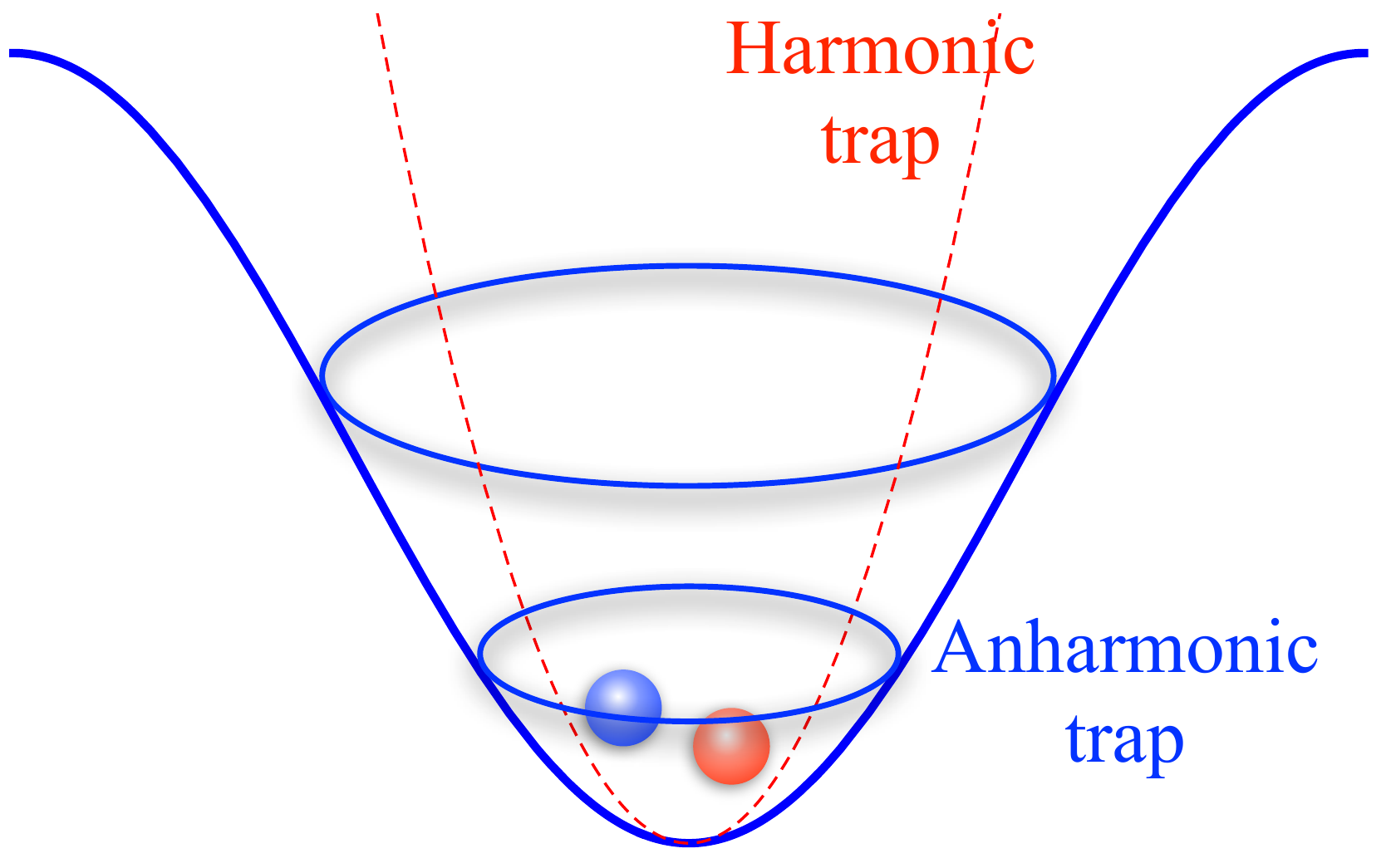}
\caption{(color online) A schematic of two ultracold atoms trapped in a confinement. Here the confinement in general is anharmonic and the two atoms can also have internal states (\textit{e.g.}, the hyperfine or Zeeman states).
}\label{fig:fig1}
\end{figure}

In this subsection we briefly introduce the two-body system we study and the basic idea of our approach. More details of our method will be shown in the following subsections.

 We consider two ultracold atoms in a confinement, as shown in Fig.~(\ref{fig:fig1}). The total Hilbert space ${\mathscr H}_{\rm tot}$ of our system is given by ${\mathscr H}_{\rm tot}={\mathscr H}_{r}\otimes{\mathscr H}_{ R}\otimes{\mathscr H}_{S}$, where ${\mathscr H}_{r}$ and ${\mathscr H}_{R}$ are the Hilbert spaces for the relative and c.m. motions, respectively, and ${\mathscr H}_{S}$ is the one of the internal space of these two atoms. In this work we denote the state in ${\mathscr H}_j$ ($j=r,R,S,{\rm tot}$) as $|\rangle_j$,
and denote the state in ${\mathscr H}_R\otimes {\mathscr H}_S$ as $|\rangle_{RS}$. Furthermore, we work in the ``${\hat {\bf r}}$-representation", with ${\hat {\bf r}}$ being the relative-position operator of these two atoms. In this representation the state $|\Psi\rangle_{\rm tot}$ of the total Hilbert space is described by the corresponding ``relative wave function"
\begin{eqnarray}
|\Psi({\bf r})\rangle_{RS}\equiv\ _{r}\!\langle{\bf r}|\Psi\rangle_{\rm tot},
\end{eqnarray}
with $|{\bf r}\rangle_r$ being the eigen-state of ${\hat {\bf r}}$. It is clear that $|\Psi({\bf r})\rangle_{RS}$ is a ${\bf r}$-dependent state in ${\mathscr H}_R\otimes {\mathscr H}_S$.

 The Hamiltonian of our system can be expressed as ($\hbar=1$)
 \begin{eqnarray}
{\hat H}&=& {\hat K}+{V}^{\rm (c)}({\hat {\bf R}},{\bf r})+{\hat h}_S(\delta)+{\hat U}_{{\rm bare}}(r),\label{h}
\end{eqnarray}
with
\begin{eqnarray}
 {\hat K}&=& -\frac{\nabla_{{\bf r}}^{2}}{2\mu}+\frac{{\hat {\bf P}}^{2}}{2M},\label{kin}
 \end{eqnarray}
where $\mu$ and $M$ are the reduced mass and total mass of the two
atoms, respectively, ${\hat {\bf R}}$ and ${\hat {\bf P}}$ are the operators of coordinate and momentum of c.m., respectively. Here ${ V}^{\rm (c)}({\hat {\bf R}},{\bf r})$ is the total confinement potential, which contains the coupling between relative and c.m. motion, and the ${\bf r}$-independent operator ${\hat h}_S$ is the free Hamiltonian of internal states ({\it e.g.}, the Zeeman energies of hyperfine states).
In realistic systems the atomic internal states are always coupled to some homogeneous external field, {\it e.g.}, a static magnetic field, and we use $\delta$ to denote the parameter of this external field.
In addition, ${\hat U}_{{\rm bare}}(r)$
is the inter-atomic interaction potential, which is a complicated function of the inter-atomic distance $r$ and satisfies ${\hat U}_{{\rm bare}}(r\rightarrow\infty)=0$.
Here we assume ${\hat U}_{{\rm bare}}(r)$ is an isotropic short-range potential with range $d_{\rm int}$, i.e., we can ignore this interaction for $r\gtrsim d_{\rm int}$. For the systems with inter-atomic van der Waals potential, we can choose $d_{\rm int}$ as the characteristic length $\beta_6$ of the van der Waals potential \cite{footnote1}.
Both ${ V}^{\rm (c)}({\hat {\bf R}},{\bf r})$ and ${\hat U}_{{\rm bare}}(r)$ may be dependent on atomic internal states. Moreover, in this work we assume that the atomic energy is low enough so that we can only consider the $s$-wave interaction.

Now we show the basic idea of our approach to solve this two-body problem. 
We notice that, in the region 
\begin{eqnarray}
r\ll d_{\rm trap} \label{sr1}
\end{eqnarray}
we can ignore the ${\bf r}$-dependence of the confinement potential, i.e., make the approximation ${ V}^{\rm (c)}({\hat {\bf R}},{\bf r})\approx { V}^{\rm (c)}({\hat {\bf R}},{\bf r=0})$. Thus, the behavior of the exact eigen-state $|\Psi_{\rm exa}({\bf r})\rangle_{RS}$ of the total Hamiltonian ${ H}$ in the region (\ref{sr1}) is approximately determined by the Schr\"odinger equation
\begin{widetext}
\begin{eqnarray}
\left[{\hat K}+{ V}^{\rm (c)}({\hat {\bf R}},{\bf r=0})+{\hat h}_S(\delta)+{\hat U}_{{\rm bare}}(r)\right] |\Psi_{\rm exa}({\bf r})\rangle_{RS}=E|\Psi_{\rm exa}({\bf r})\rangle_{RS},
\label{see}
\end{eqnarray}
and the boundary condition 
\begin{eqnarray}
|\Psi_{\rm exa}({\bf r=0})\rangle_{RS}=0.\label{bd}
\end{eqnarray}
On the other hand, as mentioned above, we consider the systems where the characteristic length $d_{\rm trap}$ of the confinement is much larger than the range $d_{\rm int}$ of the realistic interaction potential, so that there exists a short-range region
\begin{eqnarray}
d_{\rm int}\lesssim r\ll d_{\rm trap},\label{sr}
\end{eqnarray}
in which both ${\hat{U}}_{{\rm bare}}({\bf r})$ and the ${\bf r}$-dependence of the confinement potential can be ignored. Since the region of Eq.~(\ref{sr}) is a subset region of Eq.~(\ref{sr1}), the behavior of $|\Psi_{\rm exa}({\bf r})\rangle_{RS}$ in the region of Eq.~(\ref{sr}) is also determined by Eqs.~(\ref{see}) and (\ref{bd}).

The first step of our approach is to evaluate the behavior of $|\Psi_{\rm exa}({\bf r})\rangle_{RS}$ in the short-range region of Eq. (\ref{sr}) by solving Eq. (\ref{see}) with the boundary condition Eq.~(\ref{bd}).
Here we emphasize that Eq. (\ref{see}) is much easier to solve than the exact eigen-equation of ${\hat H}$, because in this equation the influence of the confinement potential to the relative motion is ignored.

After obtaining the exact wave function $|\Psi_{\rm exa}({\bf r})\rangle_{RS}$ in the region (\ref{sr}), we can construct the the scattering length operator ${\hat a}_{\rm eff}(E)$
or the multi-channel HYP
${{\hat U}_{\rm eff}(E)}={\hat a}_{\rm eff}(E)\frac{2\pi}{\mu}\delta({\bf r})\frac{\partial}{\partial r}(r\cdot)$ for our system. This HYP is required to be able to reproduce the correct behavior of the wave function in the short-range region. Explicitly, we have the following criteria for ${\hat a}_{\rm eff}(E)$:

\vspace{0.25cm}
\begin{itemize}
\item[]
{\it Criteria:}
 The solution $|\Psi_{\rm exa}({\bf r})\rangle_{RS}$ of Eq. (\ref{see})
and the solution $|{\psi}({\bf r})\rangle_{RS}$ of the equation
\begin{eqnarray}
\left[{\hat K}+{ V}^{\rm (c)}({\hat {\bf R}},{\bf r=0})+{\hat h}_S(\delta)+{\hat a}_{\rm eff}(E)\frac{2\pi}{\mu}\delta({\bf r})\frac{\partial}{\partial r}(r\cdot)\right] |{\psi}({\bf r})\rangle_{RS}=E|{\psi}({\bf r})\rangle_{RS}
\label{see2}
\end{eqnarray}
\hspace*{1.7cm}satisfy $|\Psi_{\rm exa}({\bf r})\rangle_{RS}\approx |{\psi}({\bf r})\rangle_{RS}$ for $d_{\rm int}\lesssim r\ll d_{\rm trap}$.
\end{itemize}
\vspace{0.25cm}
In the following subsection we will show the detail on how to construct the scattering length operator ${{\hat a}_{\rm eff}(E)}$. It is clear that since
Eqs. (\ref{see}) and (\ref{see2}) include
 the operators ${\hat h}_S(\delta)$ and ${ V}^{\rm (c)}({\hat {\bf R}},{\bf r=0})$, the scattering length operator
${\hat a}_{\rm eff}(E)$
would be dependent on the external-field parameter $\delta$ and the confinement potential.

When ${\hat a}_{\rm eff}(E)$ is constructed, we can calculate the scattering amplitude or energy spectrum by solving the equation
\begin{eqnarray}
\left[{\hat K}+{V}^{\rm (c)}({\hat {\bf R}},{\bf r})+{\hat h}_S(\delta)+{\hat a}_{\rm eff}(E)\frac{2\pi}{\mu}\delta({\bf r})\frac{\partial}{\partial r}(r\cdot) \right]|\Psi({\bf r})\rangle_{RS}
&=&E|\Psi({\bf r})\rangle_{RS},
\label{se2}
\end{eqnarray}
rather than the exact stationary Schr\"odinger equation ${\hat H} |\Psi_{\rm exa}({\bf r})\rangle_{RS}=E|\Psi_{\rm exa}({\bf r})\rangle_{RS}$. Namely, we replace the complicated bare interaction ${\hat U}_{{\rm bare}}(r)$ with the zero-range HYP ${{\hat U}_{\rm eff}(E)}$ corresponding to ${\hat a}_{\rm eff}(E)$, so that
the
calculation can be simplified.

The principle of this method is the same as the other zero-range effective potential, which is explained as follows. Firstly, according to our above discussion, in the short-range region 
$d_{\rm int}\lesssim r\ll d_{\rm trap}$
the solutions of Eq. (\ref{se2}) and the exact equation 
\begin{eqnarray}
{\hat H} |\Psi_{\rm exa}({\bf r})\rangle_{RS}=E|\Psi_{\rm exa}({\bf r})\rangle_{RS}\label{eeq}
\end{eqnarray}
 are approximately the same as each other. Secondly, the solutions in this region can serve as a ``boundary condition" for these two equations in the region with longer inter-atomic distance (i.e., the region with larger $r$). Thirdly, in the longer-distance these two equations also have the same form because both ${\hat U}_{\rm bare}(r)$ and the HYP can be ignored. Due to these three facts, 
the eigen-energy $E$ and the behavior of the eigenstates in the region with $r\gtrsim d_{\rm int}$, which are given by 
Eq. (\ref{se2}), would be approximately the same as the ones given by the exact equation (\ref{eeq}).

\end{widetext}

\subsection{Construction of ${{\hat a}_{\rm eff}(E)}$}
\label{sec:aeff}

Now we show the detail of our approach to construct the scattering-length operator ${{\hat a}_{\rm eff}(E)}$ for the systems with either single-component or multi-component atoms.

\subsubsection {Single-component atoms}
\label{sec:sec2B1}

We consider the system composite of two single-component atoms, where the c.m.-relative motional coupling can be induced by the
confinement potential.
For this system we only require to consider the relative and c.m. spatial motion, or the state in ${\mathscr H}_{r}\otimes{\mathscr H}_{ R}$, and the energy ${\hat h}_S$ is absent in the Hamiltonian. Furthermore,
 the relative and c.m. motion are decoupled in Eqs.~(\ref{see}) and (\ref{see2}) shown above. Due to these facts, we can construct the scattering length operator ${\hat a}_{\rm eff}(E)$ as follows.

 We first solve the single-channel Schr\"odinger equation
\begin{eqnarray}
\left\{-\frac{\nabla_{{\bf r}}^{2}}{2\mu}+{\hat U}_{{\rm bare}}(r)\right\} \psi({\bf r})=\varepsilon_{\rm r}\psi({\bf r})\label{e2}
\end{eqnarray}
for the two-atom relative motion in 3D free space, and derive the corresponding single-channel energy-dependent scattering length $a_{\rm bare}(\varepsilon_{\rm r})$, which is defined with the $s$-wave phase
shift and the wave function behavior,
as shown in Eq. (\ref{ae}) and Ref. \cite{footnote}.

Then we solve the eigen-equation for the Hamiltonian of the c.m. motion for ${\bf r=0}$, i.e., the equation
\begin{eqnarray}
\left[
\frac{{ {\bf P}}^{2}}{2M}+{ V}^{\rm (c)}({\hat {\bf R}},{\bf r=0})
\right] |{\cal E}_n\rangle_R={\cal E}_n|{\cal E}_n\rangle_R,\label{e3}
\end{eqnarray}
and derive the eigen-energies and eigen-states.

Finally,
using the above results we can construct the scattering length operator
${\hat a}_{\rm eff}(E)$ as
\begin{eqnarray}
{\hat a}_{\rm eff}(E)&=&\sum_n |{\cal E}_n\rangle_R\langle {\cal E}_n| a_{\rm bare}(E-{\cal E}_n).
\label{usp}
\end{eqnarray}
It is clear this scattering length operator satisfies the criteria in Sec.~\ref{sec:sec2A}.

\subsubsection{Multi-component atoms: simple cases}
\label{sec:sec2B2}

Now we consider the construction of ${\hat a}_{\rm eff}(E)$. In this subsection we focus on two relatively simple but quite realistic cases. The approach for the general cases will be shown in Sec. \ref{sec:sec2B4}.

{\it Simple Case 1: there is no c.m.-relative motional coupling.} We consider the system where
 the free internal-state Hamiltonian ${\hat h}_S(\delta)$ does not commute with the inter-atomic interaction ${\hat U}_{{\rm bare}}(r)$,
while
the c.m. and relative motion are not coupled with each other. For this system only the relative motion and internal state, i.e., the state in ${\mathscr H}_{r}\otimes{\mathscr H}_{S}$, are required to be considered. The scattering length operator ${\hat a}_{\rm eff}(E)$ can be constructed as follows:

We first derive the eigen-energies and eigen-states of
 ${\hat h}_S(\delta)$
 by solving the equation
\begin{eqnarray}
{\hat h}_S(\delta)|s_j\rangle_S=s_j|s_j\rangle_S;\ \ j=1,2,...N_S,
\end{eqnarray}
with $N_S$ being the dimension of ${\mathscr H}_{S}$.
Then we solve the Schr\"odinger equation
\begin{eqnarray}
\left\{-\frac{\nabla_{{\bf r}}^{2}}{2\mu}+{\hat h}_S(\delta)+{\hat U}_{{\rm bare}}(r)\right\} |\psi({\bf r})\rangle_S=E|\psi({\bf r})\rangle_S\ \ \ \ \
\label{e3a}
\end{eqnarray}
in the $s$-wave manifold for the relative motion and internal state of two atoms in 3D free space, with boundary condition $|\psi(r=0)\rangle_S=0$. This equation has $N_S$ linearly-independent solutions $|\psi^{(j)}({\bf r})\rangle_S$, ($j=1,2,...,N_S$), which satisfy the condition
\begin{eqnarray}
&&|\psi^{(j)}({\bf r})\rangle_S\nonumber\\
&=&\frac{1}{r}
\left\{
\frac{1}{k_j}\sin(k_j r)|s_j\rangle_S -
\sum_{l=1}^{N_S}
a_{lj}(E,\delta)\cos(k_{l} r)|s_{l}\rangle_S\right\}\nonumber\\
&&{\rm (for}\ \ r\gtrsim d_{\rm int}{\rm )},\label{psi12}
\end{eqnarray}
where $k_l=\sqrt{2\mu(E-s_l)}$ ($l=1,...,N_S$) with $\sqrt{z}\equiv i\sqrt{|z|}$ for $z<0$, and the parameter $a_{lj}$ depends on the energy $E$ and the external-field parameter $\delta$.
In our approach we require to derive the values of $a_{lj}(E,\delta)$ by solving Eq. (\ref{e3a}).

Here we emphasize that the calculation of $a_{lj}(E,\delta)$ ($l=1,...,N_S$) can be simplified for many realistic systems, where the bare inter-atomic interaction 
 ${\hat U}_{\rm bare}(r)$ can be approximated as a internal-state independent van der Waals potential
 beyond a particular range $b$, i.e.,
 \begin{equation}
{\hat U}_{\rm bare}\left(r>b\right)\approx-\frac{\beta_6^{4}}{2\mu r^{6}},\label{vd}
\end{equation}
with $\beta_6$ ($\beta_6>b$) being the van der Waals characteristic length and satisfying $|k_j|^2\ll1/\beta_6^2$ ($j=1,...,N_S$). For these systems one can solve Eq. (\ref{e3a}) and analytically calculate the parameters $a_{lj}(E,\delta)$ using the multi-channel quantum defect theory (MQDT), which is based on the analytical solution of the Schr\"odinger equation with the van der Waals potential \cite{Gao2007,Zhang2017,Gao2005,Gao1998QDT}. In Appendix~\ref{appqdt} we show the detail of this MQDT calculation.

After obtaining the coefficients $a_{lj}(E,\delta)$, we can construct the scattering length operator ${\hat a}_{\rm eff}(E)$ as
\begin{eqnarray}
{\hat a}_{\rm eff}(E)&=&\sum_{l,j} |s_l\rangle_S\langle s_{j}| a_{lj}(E,\delta).
\label{us}
\end{eqnarray}
Notice that ${\hat a}_{\rm eff}(E)$ depends on not only the total energy $E$, but also the external-field parameter $\delta$.
We can straightforwardly prove that ${\hat a}_{\rm eff}(E)$ satisfies the criteria shown in Sec.~\ref{sec:sec2A}.

{\it Simple case 2: ${ V}^{\rm (c)}({\hat {\bf R}},{\bf r}={\bf 0})$ is internal-state independent.}
We consider a more complicated system
with both of the two situations (A) and (B) of Sec.~\ref{sec:sec1}. Nevertheless, we assume the confinement potential in the 
short-range region
(i.e., ${ V}^{\rm (c)}({\hat {\bf R}},{\bf r}={\bf 0})$) is independent of the atomic internal state.
Thus, in Eqs. (\ref{see}) and ( \ref{see2}) for the short-range wave function, the c.m. motion is decoupled with the relative motion and the internal state, as in the case of Sec.~\ref{sec:sec2B1}. Therefore, we can construct ${\hat a}_{\rm eff}(E)$ by cominbing the techniques in the above two cases. Explicitly,
we first construct a scattering length operator only for the relative motion via the method of the {\it simple case 1} of this subsection, and then take into account the
 c.m. motion using the eigen-value ${\cal E}_n$ and the eigen-state $|{\cal E}_n\rangle_R$ of the Hamiltonian $\hat {\bf P}^{2}/(2M)+{ V}^{\rm (c)}({\hat {\bf R}},{\bf r=0})$, as in Sec.~\ref{sec:sec2B1}. The scattering length operator ${\hat a}_{\rm eff}(E)$, which satisfies the criteria of Sec.~\ref{sec:sec2A}, can be expressed as
\begin{eqnarray}
{\hat a}_{\rm eff}(E)&=&\sum_n |{\cal E}_n\rangle_R\langle {\cal E}_n| \otimes
\left[\sum_{l,j} |s_l\rangle_S\langle s_{j}| a_{lj}(E-{\cal E}_n,\delta)\right],\nonumber\\
\end{eqnarray}
where the functions $a_{lj}(E,\delta)$ ($l,j=1,...,N_S$) are defined in
 Eq. (\ref{psi12}), and can be derived from the relative Schr\"odinger equation Eq. (\ref{e3a}), as shown in the {\it simple case 1} of this subsection.

\subsubsection{Multi-component atoms: general cases}
\label{sec:sec2B4}

For the most general cases of multi-component atoms with both situations (A) and (B), we can construct the scattering length operator ${\hat a}_{\rm eff}(E)$ by directly generalizing the method of the above subsections. Explicitly, we first derive the eigen-states and eigen-energies of the Hamiltonian
${{\hat {\bf P}}^{2}}/{(2M)}+{ V}^{\rm (c)}({\hat {\bf R}},{\bf r=0})+{\hat h}_S(\delta)$ for the c.m. motion and internal state by solving
\begin{eqnarray}
&&\left[\frac{{\hat {\bf P}}^{2}}{2M}+{ V}^{\rm (c)}({\hat {\bf R}},{\bf r=0})+{\hat h}_S(\delta)\right]|\lambda_n\rangle_{RS}=\lambda_n|\lambda_n\rangle_{RS}\nonumber\\
&&(n=1,...,N_{RS}),
\label{nn}
\end{eqnarray}
with $N_{RS}$ being the dimension of ${\mathscr H}_{R}\otimes{\mathscr H}_{S}$. Then we
directly solve Eq. (\ref{see}) under the boundary condition $|\Psi_{\rm exa}({\bf r=0})\rangle_{RS}=0$ in the $s$-wave manifold, and derive the linearly-independent special solutions $|\Psi_{\rm exa}^{(j)}(r)\rangle_{RS}$, which satisfy
\begin{eqnarray}
&&|\Psi_{\rm exact}^{(j)}(r)\rangle_{RS}\nonumber\\
&=&\frac{1}{r}
\left\{
\frac{1}{p_j}\sin(p_j r)|\lambda_j\rangle_{RS} -
\sum_{l=1}^{N_{RS}}
A_{lj}(E,\delta)
\cos(p_{l} r)|\lambda_{l}\rangle_{RS}\right\}\nonumber\\
&&{\rm (for}\ \ r\gtrsim d_{\rm int}{\rm )},\label{psi12b}
\end{eqnarray}
with $p_l=\sqrt{2\mu(E-\lambda_l)}$ ($l=1,...,N_{RS}$). Similar as in Sec.~\ref{sec:sec2B2}, if ${\hat U}_{\rm bare}(r)$ can be approximated as an internal-state independent van der Waals potential beyond a critical range, i.e., satisfies the condition (\ref{vd}) as well as $|p_j|^2\ll1/\beta_6^2$ ($j=1,...,N_{RS}$), the coefficients $A_{lj}(E,\delta)$ ($l,j=1,...,N_{RS}$) can be obtained with MQDT, as shown in Appendix \ref{appA2}.
Finally, the scattering length operator ${\hat a}_{\rm eff}(E)$ satisfying the criteria in Sec.~\ref{sec:sec2A}, can be expressed in terms of $A_{lj}(E,\delta)$ as
\begin{eqnarray}
{\hat a}_{\rm eff}(E)&=&\sum_{l,j} |\lambda_l\rangle_{RS}\langle \lambda_{j}| A_{lj}(E,\delta).
\label{us22}
\end{eqnarray}

\section{Two alkaline-earth (like) atoms in an optical lattice site}
\label{sec:sec3}

In Sec.~\ref{sec:sec2} we have shown our HYP approach for the two-body problem of
confined ultracold atoms. As a demonstration, here we calculate
the energy spectrum of 
two confined
alkaline-earth (like) atoms. We introduce the
properties of this system and our calculation method
in this section, and 
 compare our theoretical
results with the recent experiments of ultracold $^{173}{\rm {Yb}}$
atoms and $^{171}{\rm {Yb}}$ atoms \cite{Scazza_2014,PRL2015,PRX2019,PRL2014,PRA2019,Bettermann,Benjamin} in Sec.~\ref{sec:sec4}.

\begin{figure}
\includegraphics[scale=0.45]{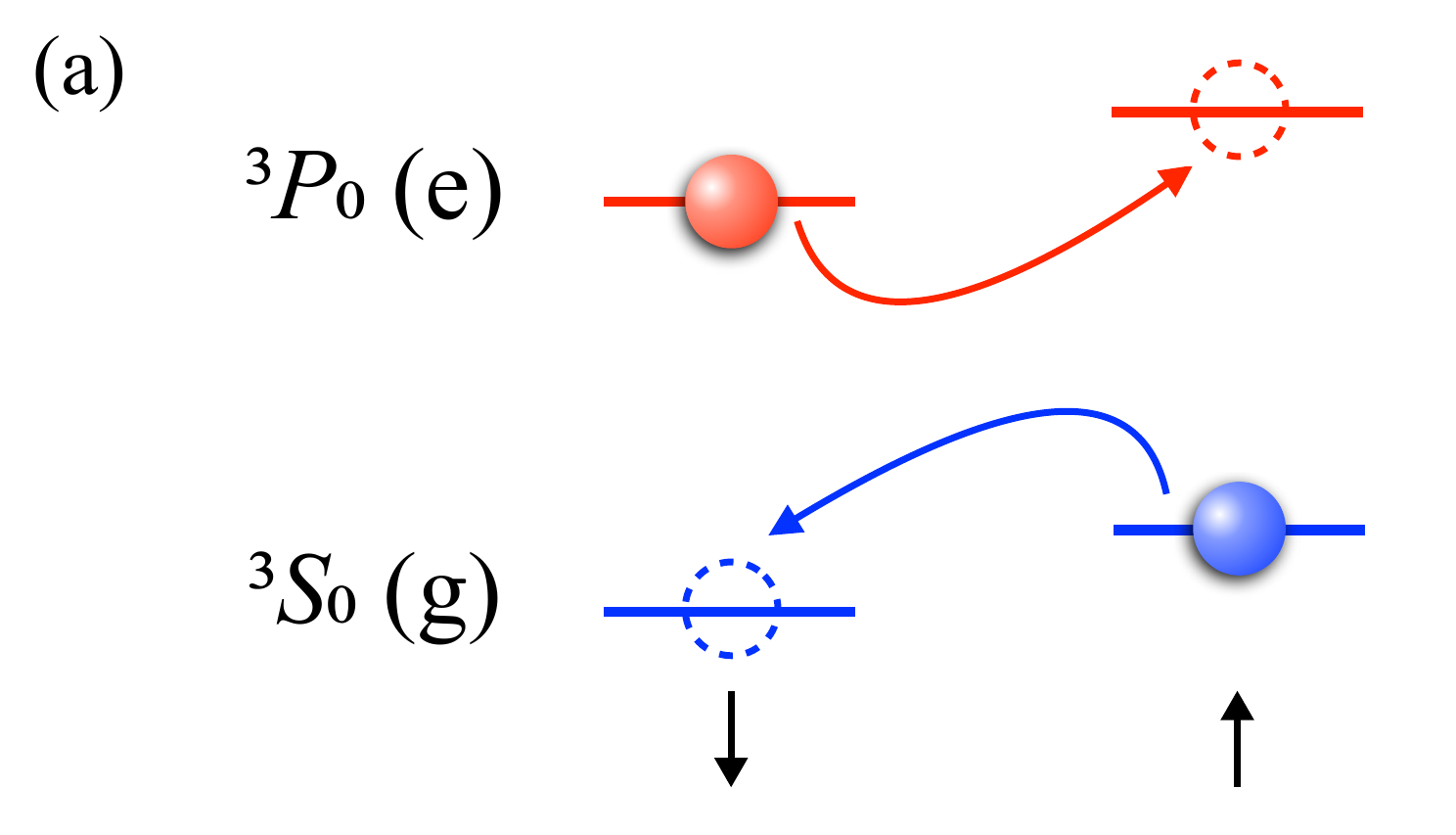}
\includegraphics[scale=0.4]{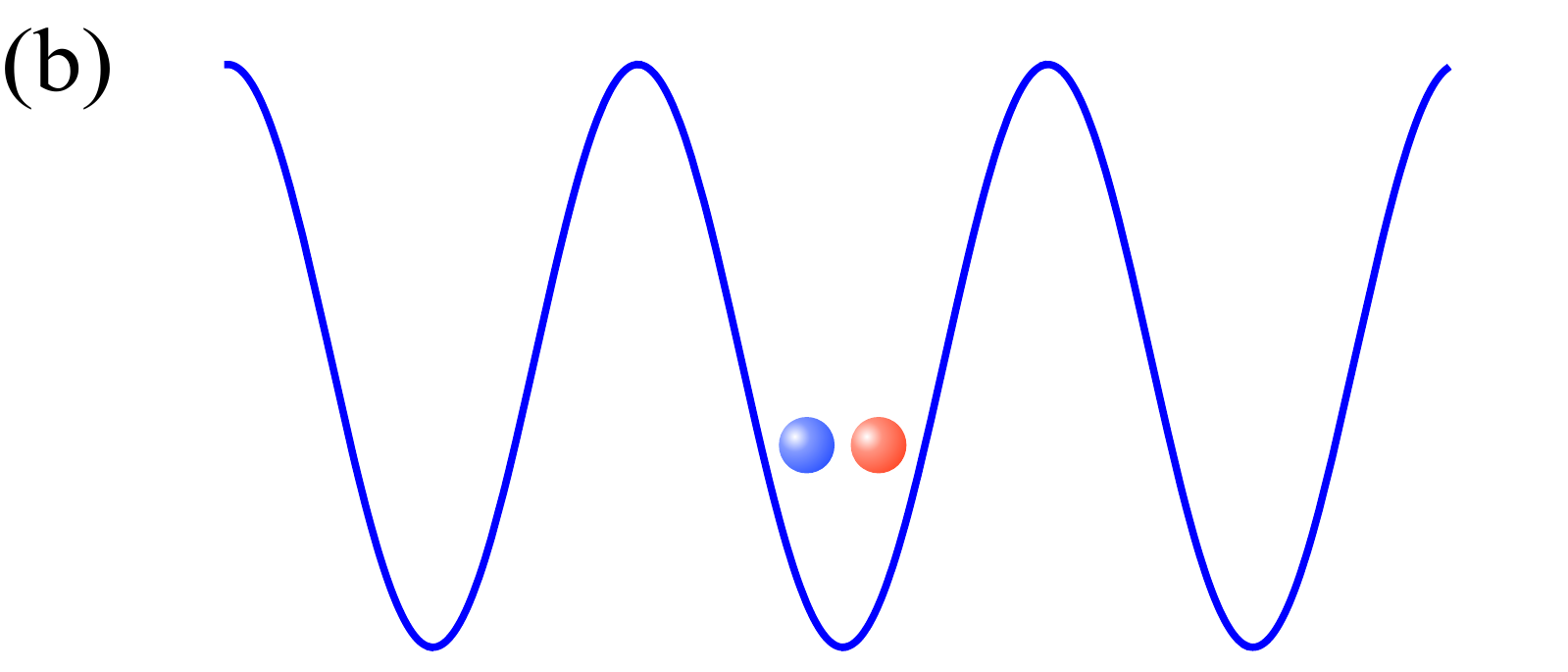}
\includegraphics[scale=0.35]{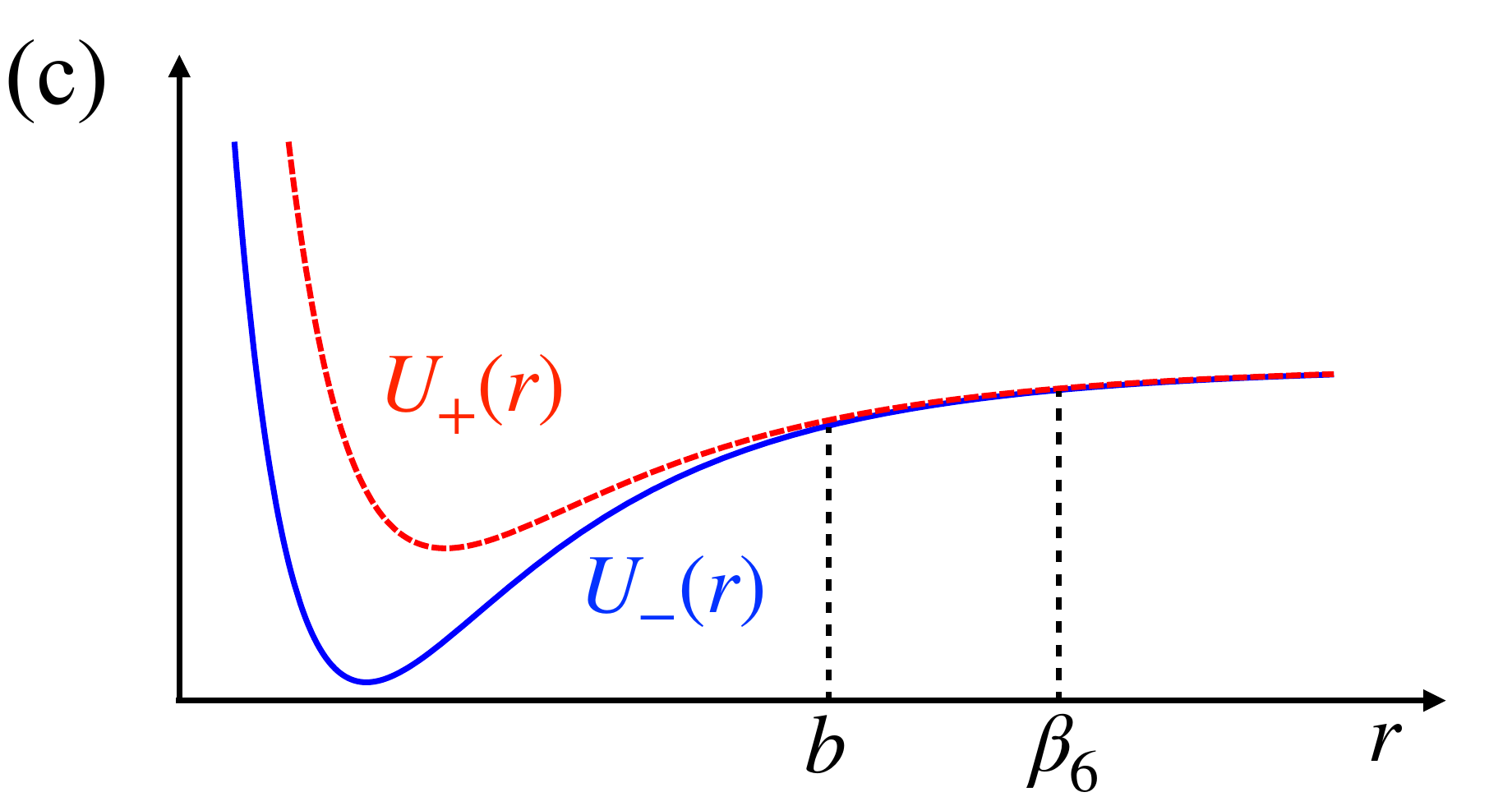}
\caption{(color online) {\bf (a):} Energy levels of the alkaline-earth atoms. The $e$- and $g$-atoms can exchange their nuclear-spin states during inter-atomic collision. {\bf (b):} Schematic of the system with two ultracold Yb atoms confined in a site of a 3D optical lattice, which are in $e$- (red) and $g$- (blue) states, respectively. {\bf (c):} The bare inter-atomic interaction potential curve $U_\pm(r)$ of two Yb atoms in anti-symmetric and symmetric nuclear-spin states $| \pm \rangle_S$. Beyond a particular range $b$, the inter-atomic interaction can be approximated as an internal-state independent van der Waals potential with characteristic length $\beta_6$.}\label{fig:fig2}
\end{figure}

\subsection{Properties of atom and confinement}
\label{sec:sec3A}

{\begin{figure*}
\includegraphics[scale=0.4]{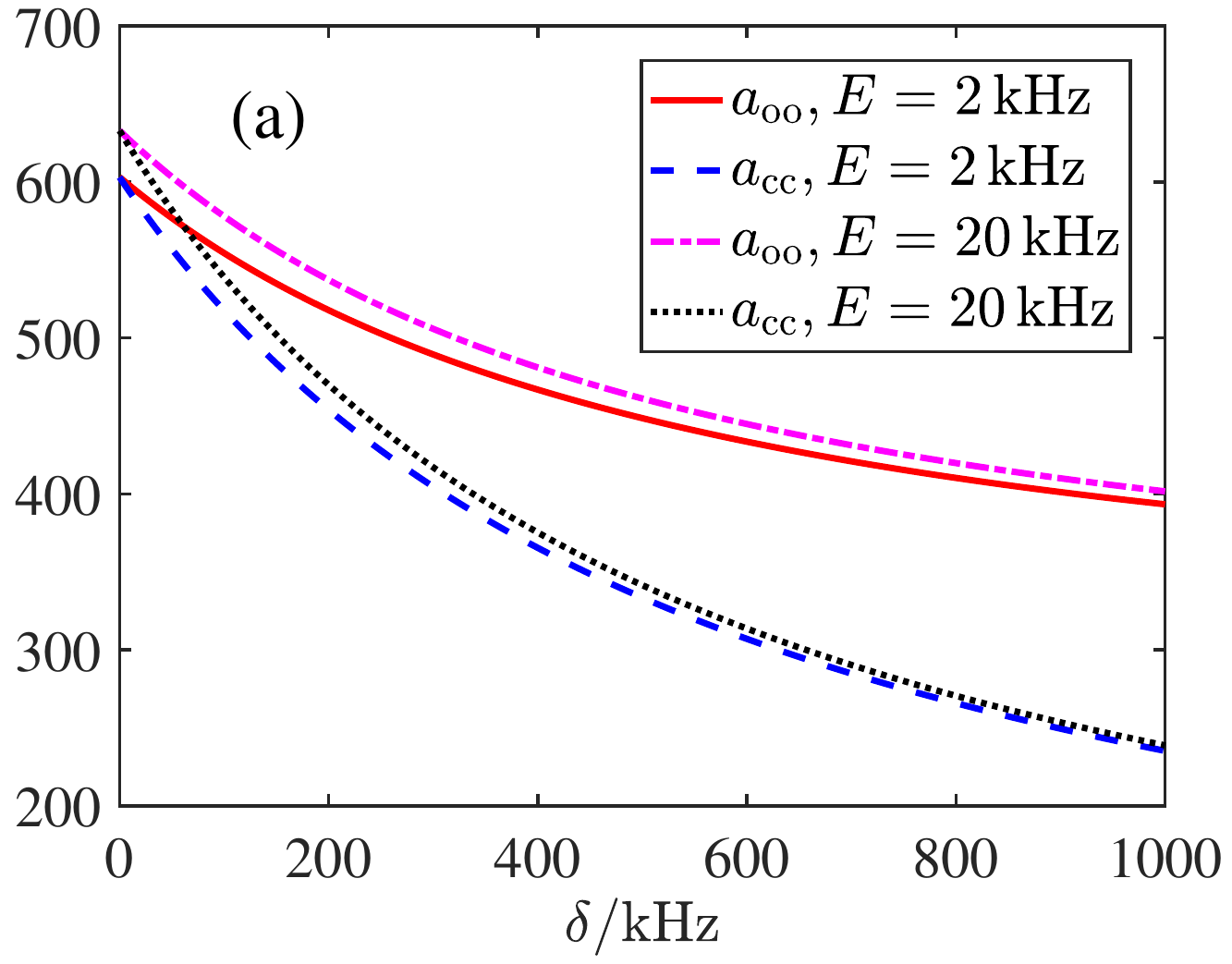}
\includegraphics[scale=0.4]{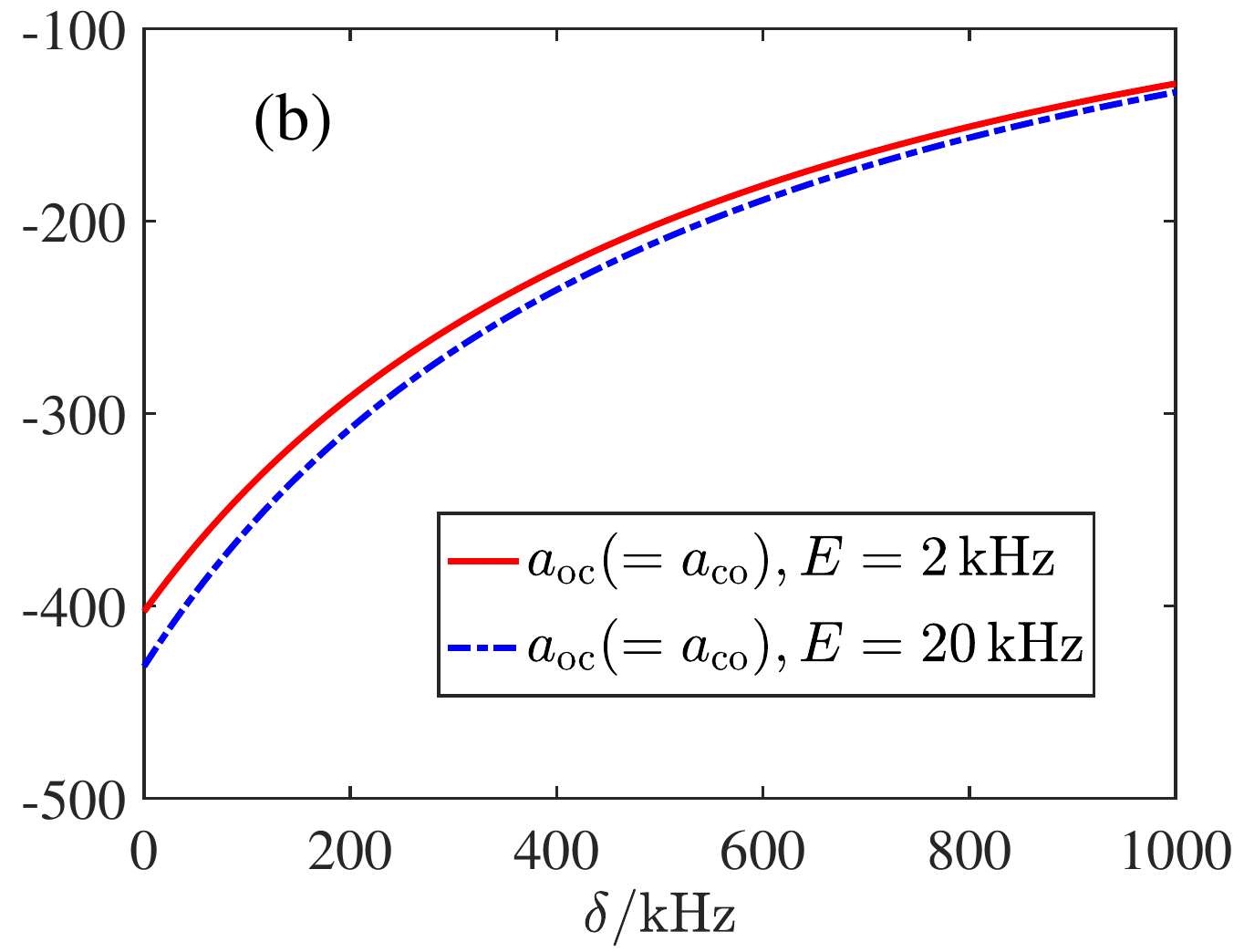}
\includegraphics[scale=0.4]{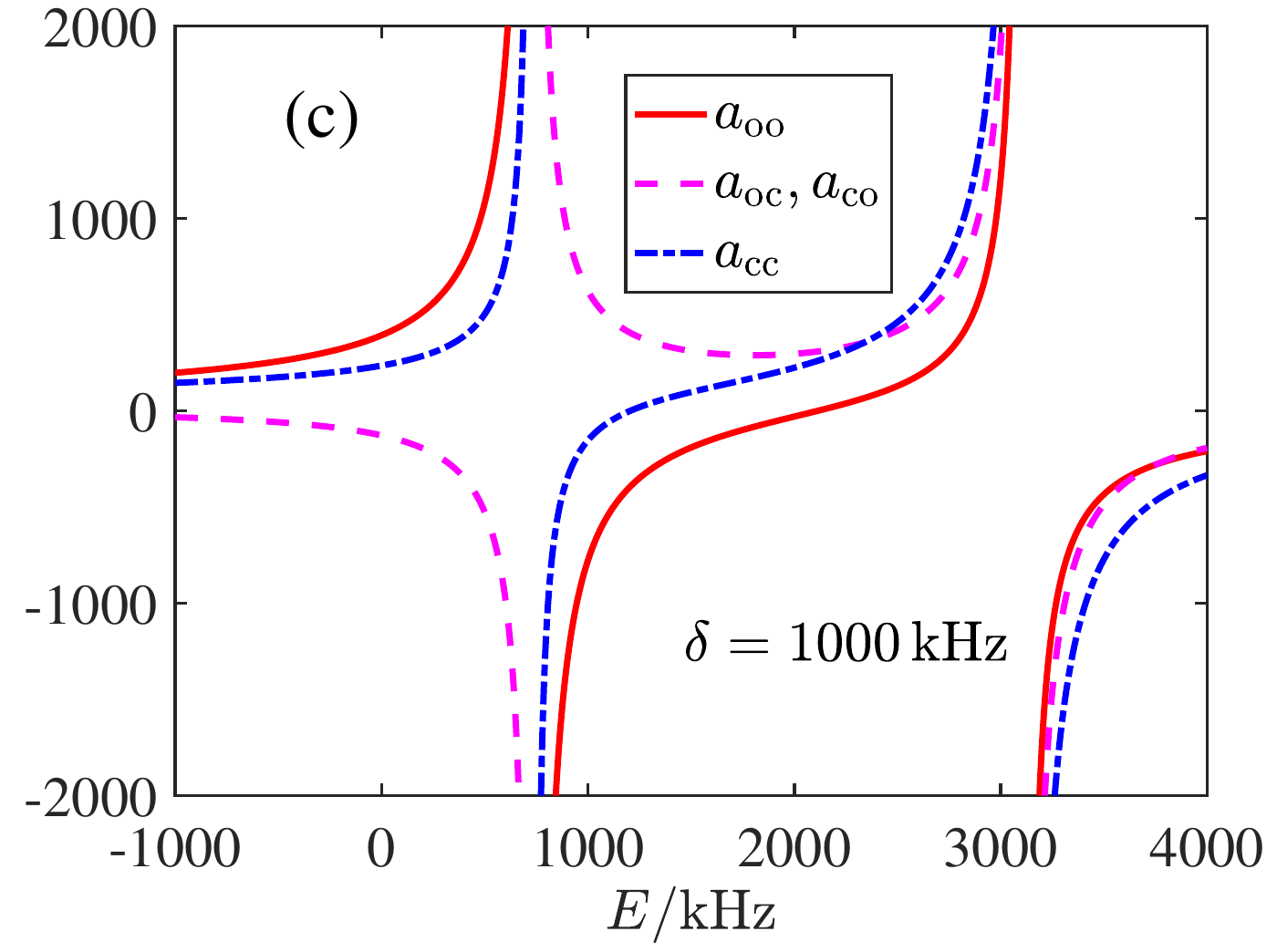}
\caption{(color online) The coefficients $a_{i,j}$ ($i,j=o,c$) 
typical parameters $\mu=100m_p$, $a_+ =1000a_0, a_- = 200a_0$, $\beta_6 = 150a_0$, with $m_{\rm p}$ and $a_0$ being the mass of a proton and Bohr's radius, respectively. Here we show $a_{i,j}$ ($i,j=o,c$) 
 as a function of the Zeeman gap $\delta$ {\bf ((a) and (b))} and the collision energy $E$ {\bf (c)} \cite{footnote4}. For our problem we always have $a_{oc}=a_{co}$.
}
\label{qdt}
\end{figure*}

As shown in Fig.~(\ref{fig:fig2}a), we consider two homonuclear
fermionic alkaline-earth (like) atoms, which are in 
electronic $^{1}{\rm S}_{0}$ ($g$) and $^{3}{\rm P}_{0}$ ($e$)
states, respectively. Mathematically we can treat the $e$ atom and $g$ atom as two distinguishable atoms. We further assume that each atom can be in nuclear-spin state $\uparrow$ or $\downarrow$, corresponding to different magnetic quantum numbers, and the nuclear-spin states of the two atoms are different. Explicitly, for our two-body system the Hilbert space ${\mathscr H}_{S}$ of the two-body internal state is spanned by the two states:
\begin{eqnarray}
 \vert c\rangle_S\equiv\vert\downarrow\rangle_{e}\vert\uparrow\rangle_{g}
 \ \ \ {\rm and}
 \ \ \
 \vert o\rangle_S\equiv\vert\uparrow\rangle_{e}\vert\downarrow\rangle_{g}.
\end{eqnarray}
 In the
presence of the bias magnetic field, the Land\'e $g$-factor of the $e$  and the $g$ state is different \cite{YeJunPRA}. As a result, the states $|o\rangle_S$ and $|c\rangle_S$ have different Zeeman energies. Thus, for our system the free internal-state
 Hamiltonian ${\hat h}_S(\delta)$ defined in Sec.~\ref{sec:sec2} can be expressed as
 \begin{eqnarray}
 {\hat h}_S(\delta)=\delta|c\rangle_S\langle c|.\label{hhs}
 \end{eqnarray}
Here we have chosen the Zeeman-energy difference between the two internal states, which is proportional to the magnetic field, as the external-field parameter $\delta$.

Moreover, as shown in Fig.~(\ref{fig:fig2}b), we assume the atoms are confined in a site of a 3D optical lattice formed by lasers with magnetic wave length, so that the two atoms experience the same confinement potential. This potential is also independent of the nuclear-spin state and can be expressed as
\begin{eqnarray}
V_{{\rm {opt}}}({\bf r}^{(j)})=\frac{s_{f}k_{L}^{2}}{2m}\sum_{\alpha=x,y,z}\sin^{2}[k_{L}r^{(j)}_{\alpha}],\ \ (j=g,e),\nonumber\\
\label{eq:optV}
\end{eqnarray}
where $m=2\mu$ is the single-atom mass, ${\bf r}^{(j)}\equiv(r^{(j)}_{x},r^{(j)}_{y},r^{(j)}_{z})$ ($j=g,e$) is the coordinate of the $j$-atom,
 and $k_{L}$ and $s_{f}$ are the wavenumber and the
dimensionless lattice depth, respectively.
In our calculation we expand
this potential around the minimum point ${\bf r}^{(j)}={\bf 0}$ and keep the terms up to $|{\bf r}^{(j)}|^6$.

As in Sec.~\ref{sec:sec2}, in further calculations we express the total confinement potential $V^{\rm (c)}=V_{{\rm {opt}}}({\bf r}^{(e)})+V_{{\rm {opt}}}({\bf r}^{(g)})$ as a function of the relative position ${\bf r}={\bf r}^{(e)}-{\bf r}^{(g)}$ and
the c.m. position operator $ {\hat {\bf R}}=({\bf r}^{(e)}+{\bf r}^{(g)})/2$. The straightforward calculation yields
\begin{eqnarray}
 V^{\rm (c)}( {\hat {\bf R}}, {\bf r})=V_{0r}^{\rm (c)}({\bf r})+V_{0R}^{\rm (c)}({\hat {\bf R}})+V_{1}^{\rm (c)}({\hat {\bf R}},{\bf r}),\label{vvc}
\end{eqnarray}
with the terms in the right-hand side being defined as
\begin{eqnarray}
V_{0r}^{\rm (c)}({\bf r})&=&\frac{1}{2}\mu\omega^{2}r^{2},\\
\nonumber\\
V_{0R}^{\rm (c)}( {\hat {\bf R}})&=&2\mu\omega^{2}|{\hat {\bf R}}|^{2}
+2\sum_{\alpha=x,y,z}\left[\xi_{4}\hat{ R}_\alpha^{4}+\xi_{6}\hat{ R}_\alpha^{6}\right],\label{V0R}
\end{eqnarray}
and
\begin{eqnarray}
V_{1}^{\rm (c)}({\hat {\bf R}}, {\bf r})&=&\sum_{\alpha=x,y,z}\left[ \frac{1}{8}\xi_{4} {r}_\alpha^{4}+\frac{1}{32}\xi_{6} {r}_\alpha^{6}+3\xi_4 r_\alpha^2 \hat{R}_\alpha^2 \right.\nonumber\\
&&\qquad\quad +\left. \frac{15}{2}\xi_6 r_\alpha^2 \hat{R}_\alpha^4 + \frac{15}{8}\xi_6 r_\alpha^4 \hat{R}_\alpha^2\right],\label{Hpert}
\end{eqnarray}
respectively, where ${\bf r}=(r_x,r_y,r_z)$, ${\hat {\bf R}}=(\hat{ R}_x, \hat{ R}_y, \hat{ R}_z)$, and the frequency $\omega$ and the parameters $\xi_{4,6}$ are given by
\begin{eqnarray}
\omega=\frac{\sqrt{s_{f}} k_{L}^{2}}{m},\ \xi_{4}=-\frac{m^{2}\omega^{3}}{6\sqrt{s_{f}}},\ \xi_{6}=\frac{m^{3}\omega^{4}}{45s_{f}}.\label{conff}
\end{eqnarray}

Furthermore, the bare inter-atomic interaction $\hat{ U}_{\rm bare}(r)$ between these two atoms is diagonal in the internal-state basis \cite{YeJunScience,PRL2014,Scazza_2014}
\begin{eqnarray}
\vert\pm\rangle_S=\frac{1}{\sqrt{2}}\left(\vert c\rangle_S\mp\vert o\rangle_S\right),
\end{eqnarray}
and can be expressed as
\begin{eqnarray}
{\hat{U}}_{\rm bare}(r)=U_+(r)|+\rangle_S\langle +|+U_-(r)|-\rangle_S\langle -|.\label{uub}
\end{eqnarray}
with $U_{\pm}(r)$ being the interaction potential curves corresponding to states $|\pm\rangle_S$. For our system $U_{\pm}(r)$ have the same van der Waals characteristic length $\beta_6$, i.e., ${\hat{U}}_{\rm bare}(r)$ satisfies the condition (\ref{vd}), as shown in Fig.~(\ref{fig:fig2}c).

\subsection{$(E,\delta)$-dependent scattering length operator}
\label{sec:sec3B}

 To calculate the energy spectrum for our system, we first construct the scattering length operator ${\hat a}_{\rm eff}(E)$
 with the approach shown in Sec.~\ref{sec:sec2}.
 According to the above section, the total Hamiltonian $\hat{H}$ of the two alkaline-earth (like) atoms is given by Eq. (\ref{h}), with the free internal-state Hamiltonian ${\hat h}_S(\delta)$, the confinement potential ${ V}^{\rm (c)}({\hat {\bf R}}, {\bf r})$, and the
bare inter-atomic interaction ${\hat U}_{{\rm bare}}(r)$ being given by Eq. (\ref{hhs}), Eq. (\ref{vvc}), and Eq. (\ref{uub}), respectively. Since ${ V}^{\rm (c)}({\hat {\bf R}},{\bf r}={\bf 0})={ V}_{0R}^{\rm (c)}({\hat {\bf R}})$ is independent of atomic internal state, our system is in the {\it simple case 2} of Sec.~\ref{sec:sec2B2}. Using the method of that subsection, we derive the scattering length operator ${\hat a}_{\rm eff}(E)$:
\begin{eqnarray}
{\hat a}_{\rm eff}(E)=\sum_n |{\cal E}_n\rangle_R\langle {\cal E}_n| \otimes \left[ \sum_{l,j=o,c} |l\rangle_S\langle j| a_{lj}(E-{\cal E}_n,\delta)\right].\nonumber\\
\label{uae2}
\end{eqnarray}
Here ${\cal E}_n$ and $|{\cal E}_n\rangle_R$ ($n=1,2,...$) are the eigen-value and eigen-states of the c.m. Hamiltonian
\begin{eqnarray}
{\hat H}_R\equiv \frac{{ {\bf \hat{P}}}^{2}}{2M}+V_{0R}^{\rm (c)}({\hat {\bf R}}),\label{hr}
\end{eqnarray}
and can be derived by numerical diagonalization of ${\hat H}_R$.
In addition, as mentioned in the {\it simple case 2} of Sec.~\ref{sec:sec2B2}, the coefficients $a_{lj}(E,\delta)$ $(l,j=o,c)$ in the expression (\ref{uae2}) of ${\hat a}_{\rm eff}(E)$ are determined by the relative Schr\"odinger equation Eq. (\ref{e3a}) for the cases only with the free internal-state Hamiltonian ${\hat h}_S(\delta)$ and without the confinement potential ${V}^{\rm (c)}({\hat {\bf R}},{\bf r})$, and can be derived via the MQDT approach shown in Appendix \ref{appqdt}. In this appendix we show the analytical expressions of $a_{lj}(E,\delta)$ $(l,j=o,c)$  for our system, which depend on not only the energy $E$ and the Zeeman energy gap $\delta$, but also 
the van der Waals characteristic length $\beta_6$ as well as the zero-energy scattering lengths $a_+$ and $a_-$ corresponding to each potential curve $U_{+}(r)$ and $U_{-}(r)$, respectively.
In Fig.~\ref{qdt} we illustrate the coefficients $a_{lj}(E,\delta)$ $(l,j=o,c)$ for a group of typical parameters.

\subsection{Iterative calculation of energy spectrum}
\label{sec:sec3C}

Using the HYP with the scattering-length operator ${\hat a}_{\rm eff}(E)$ derived above, we can calculate the energy spectrum for the two alkaline-earth (like) atoms. To this end, we solve the Schr\"odinger equation
\begin{eqnarray}
 {\hat H}_{\rm eff}(E_b)|\Psi(r)\rangle_{RS}=E_b\,|\Psi(r)\rangle_{RS},\label{err}
\end{eqnarray}
under the boundary condition $|\Psi(r\rightarrow\infty)\rangle_{RS}=0$, with the effective Hamiltonian 
\begin{eqnarray}
&& {\hat H}_{\rm eff}(E_b)\nonumber\\
&&\equiv {\hat K}+V^{\rm (c)}( {\hat {\bf R}},{\bf r})+\delta\,\vert c\rangle_S\langle c\vert+{\hat a}_{\rm eff}(E_b)\frac{2\pi}{\mu}\delta({\bf r})\frac{\partial}{\partial r}(r\cdot).\nonumber\\
\label{heff}
\end{eqnarray}
Here the scattering length operator ${\hat a}_{\rm eff}(E_b)$ is given by Eq. (\ref{uae2}), and the terms ${\hat K}$ and $V^{\rm (c)}({\hat {\bf R}}, {\bf r})$ are given by Eq. (\ref{kin}) and Eq. (\ref{vvc}), respectively.

Since the to-be-calculated eigen-energy $E_b$ appears in both sides of Eq. (\ref{err}), we solve this equation self-consistently with an iterative approach.
We can explain this approach by taking as an example the calculation of the
ground energy $E_{g}$ (Fig.~\ref{fig:fs}). In the zero-th order calculation, we ignore the potential ${ V}_1^{\rm (c)}$, which only includes high order terms of the distance between the atoms and the trap center. Explicitly, we solve equation
\begin{eqnarray}
 {\hat H}^\prime_{\rm eff}(E_g^{(0)})|\Psi(r)\rangle_{RS}=E_g^{(0)}|\Psi(r)\rangle_{RS},\label{err2}
 \end{eqnarray}
with
\begin{eqnarray}
{\hat H}^\prime_{\rm eff}(E)\equiv
 {\hat H}_{\rm eff}(E)-{ V}_1^{\rm (c)}({\hat {\bf R}},{\bf r}),
 \label{err2a}
\end{eqnarray}
to derive the zero-th order result $E_g^{(0)}$ of the ground-state energy. Since ${\hat H}^\prime_{\rm eff}$ does not include the coupling between the c.m. and relative motions, we can solve Eq. (\ref{err2}) by separating these two degree of freedoms and straightforwardly generalizing the seminal work of T. Bush \cite{Busch1998}, with the details being shown in Appendix \ref{sec:appB}. 

\begin{figure}[t]
\includegraphics[scale=0.8]{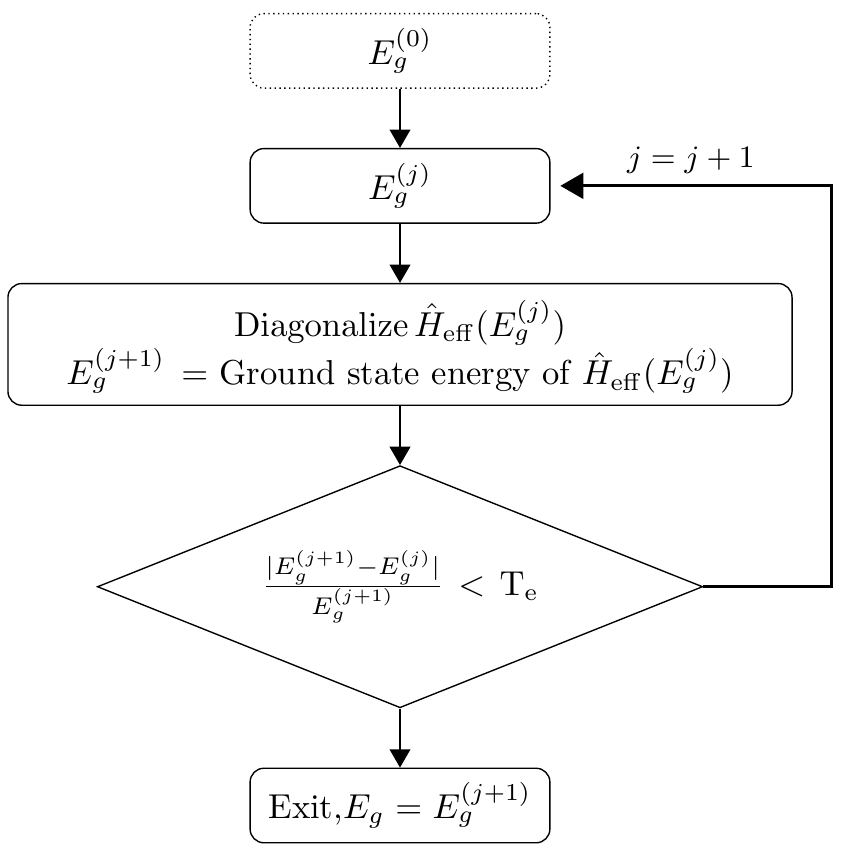} \caption{The flowchart of the iterative calculation for the ground state $E_g$. The details are explained in Sec. \ref{sec:sec3C}.
}
\label{fig:fs}
\end{figure}

Then we use $E_g^{(0)}$ as the input parameter of the first
iterative cycle, and diagonalize the Hamiltonian $\hat{H}_{\rm eff}$ with argument $E_g^{(0)}$, i.e, the Hamiltonian $\hat{H}_{\rm eff}(E_g^{(0)})$. 
Some details on our method for the diagonalization of this Hamiltonian are explained in Appendix \ref{sec:appC}. The ground state energy of $\hat{H}_{\rm eff}(E_g^{(0)})$ which is denoted as $E_{g}^{(1)}$, is the result of this cycle. Similarly, in the second iterative cycle we diagonalize the Hamiltonian $\hat{H}_{\rm eff}(E_g^{(1)})$, also with the method shown in Appendix \ref{sec:appC}, and the ground state energy $E_{g}^{(2)}$ is derived as the second-cycle result.

\begin{figure*}
\includegraphics[scale=0.5]{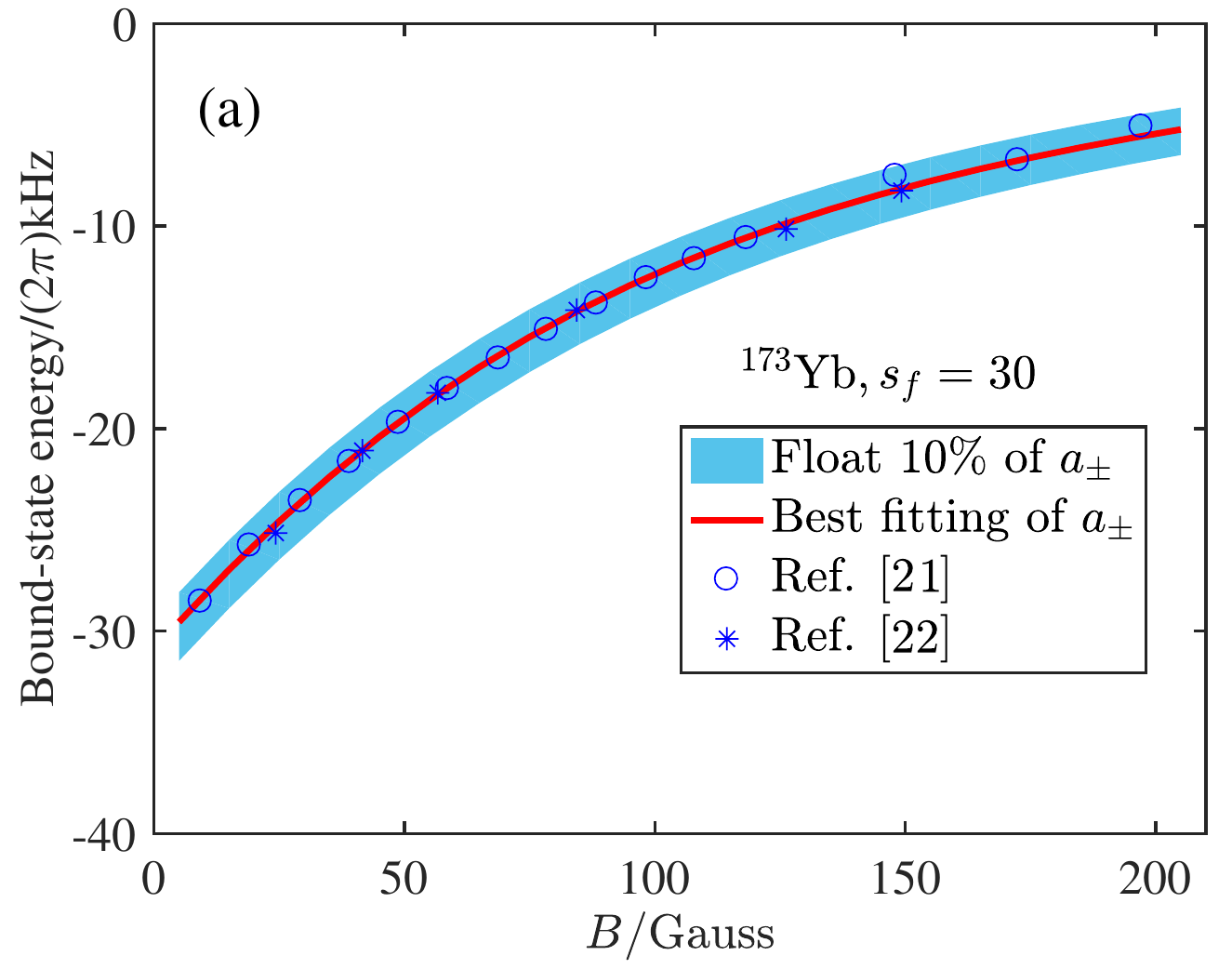} 
\includegraphics[scale=0.5]{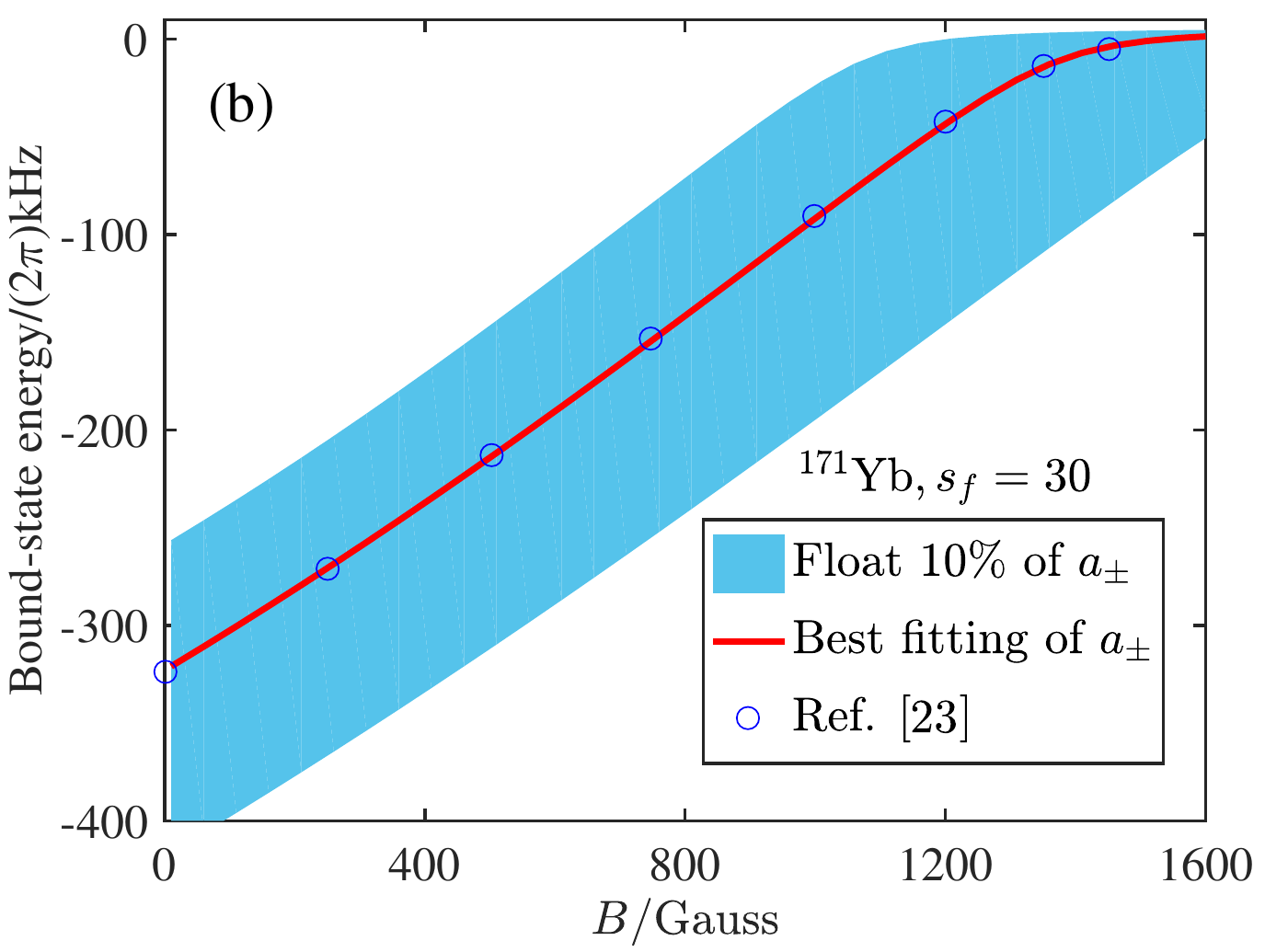} 
\caption{(color online) The bound-state energies of shallowest bound states of two ultracold $^{173}{\rm {Yb}}$ atoms {\bf (a)} and two $^{171}{\rm {Yb}}$ atoms {\bf (b)} as functions of the bias magnetic field $B$ by taking $E_g^{\rm (c.m.)}=0$, where $E_g^{\rm (c.m.)}=0$ is the two-body ground state energy without the interatomic interactions. Here we show  
 the results given by our calculations with the method in Sec. \ref{sec:sec3} (solid lines) for the best-fitting parameters $a_{+}=2012 a_{0}$, $a_{-}=193 a_{0}$ ($^{173}{\rm {Yb}}$) and $a_{+}=232 a_{0}$, $a_{-}=372 a_{0}$ ($^{171}{\rm {Yb}}$), as well as the experimental results given by Refs. \cite{PRL2015,PRX2019} ($^{173}{\rm {Yb}}$) and Ref. \cite{Bettermann} ($^{171}{\rm {Yb}}$) (stars and open circles). The blue region is the range of $\pm10\%$ variation of $a_{\pm}$.
 }
\label{fig:YbFit}
\end{figure*}

As shown in Fig.~\ref{fig:fs}, the iterative process repeats
until a tolerance requirement $|E_g^{(j+1)}-E_g^{(j)}|/|E_g^{(j+1)}|<{\rm T_e}$ is satisfied, with ${\rm T_e}$ being a threshold of relative error, which is taken as $10^{-6}$ in our calculation. When this requirement is satisfied by the $j$-th and $(j+1)$-th results $E_g^{(j)}$ and $E_g^{(j+1)}$, we suppose that the results of our calculations approximately converges to $E_g^{(j+1)}$, and thus take $E_g^{(j+1)}$ as the derived ground-state energy $E_g$ of these two atoms.

\section{Calibration of $a_{\pm}$ for $^{173}{\rm {Yb}}$ and $^{171}{\rm {Yb}}$ }
\label{sec:sec4}

In the above section we show our approach for the calculation of the eigen-energies of two alkaline-earth (like) atoms 
in the system described in Sec. \ref{sec:sec3A}. This two-body system has been realized in various experiments 
for $^{173}{\rm {Yb}}$ atoms ~\cite{PRL2015,PRX2019} 
or $^{171}{\rm {Yb}}$ atoms~\cite{Bettermann}.
In these experiments the optical lattice is prepared with lasers with magic wavelength $\lambda_{L}\equiv2\pi/k_L=759.3\,{\rm {nm}}$, so that the $g$- and $e$-atoms experience the same trapping potential given in Eq. (\ref{eq:optV})
(Fig.~\ref{fig:fig2}b).
In addition, the Zeeman-energy gap $\delta$ in Eq. (\ref{hhs}) is given by $\delta= 2\pi(m_\downarrow-m_\uparrow)\mu_B\Delta g B$,
 where $B$ is the bias magnetic field, $m_{\downarrow(\uparrow)}$ is the magnetic quantum number of nuclear-spin state $\downarrow (\uparrow)$, $\mu_B$ is the Bohr's magnetic moment, and $\Delta g$ is the difference between the Land\'e $g$-factors of the $e$- and $g$-states. Explicitly, we have $m_\downarrow=5/2$, $m_\uparrow=-5/2$, $\Delta g = 112\,{\rm Hz/G}$ for $^{173}\rm{Yb}$ atoms in the experiments of Refs. \cite{PRL2015,PRX2019}, and $m_\downarrow=-1/2$, $m_\uparrow=1/2$, $\Delta g = - 400\,{\rm Hz/G}$ for $^{171}\rm{Yb}$ atoms in the experiments of Ref. \cite{Bettermann}. 
In these experiments the two-atom eigen-energies can be measured
as a function of $B$
 via the optical absorption spectrum. 

On the other hand, as shown in Sec. \ref{sec:sec3} and Appendix \ref{sec:appA}, these bound-state energies are determined by the zero-energy scattering lengths $a_{\pm}$ with respect to the interaction potential $U_{\pm}(r)$ defined in Eq. (\ref{uub}), corresponding to the two-atom internal states $|\pm\rangle_S$.
Therefore, we can extract the values of $a_{\pm}$ for $^{173}{\rm {Yb}}$ or $^{171}{\rm {Yb}}$ atoms by fitting the eigen-energies calculated via our method shown in Sec. \ref{sec:sec3} with these experimental measurements. 

In this work we perform such fitting for the bound-state energy $E_{\rm sb}$ of the shallowest bound states of two $^{173}{\rm {Yb}}$ atoms and two $^{171}{\rm {Yb}}$ atoms in the lattices with dimensionless depth $s_f=30$, which were measured in Refs. \cite{PRL2015,PRX2019} and Ref. \cite{Bettermann}, respectively. 
We find that the best fitting parameters are $a_{+}=2012(19)\,a_{0},a_{-}=193(4)\,a_{0}$ for $^{173}{\rm {Yb}}$ atoms, and $a_{+}=232(3)\,a_{0},a_{-}=372(1)\,a_{0}$ for $^{171}{\rm {Yb}}$ atoms. Here we use the nonlinear least square fitting method \cite{NLF}, and our method for the estimation of uncertainty are shown in the Appendix \ref{sec:appD}. In our calculation the van der Waals characteristic length $\beta_6$ is taken to be $\beta_6 = 168.6\,a_{0}$ for $^{173}{\rm {Yb}}$ atoms and $\beta_6 = 168.1\,a_{0}$ for $^{171}{\rm {Yb}}$ atoms \cite{Porsev}.

\begin{figure*}
\includegraphics[scale=0.4]{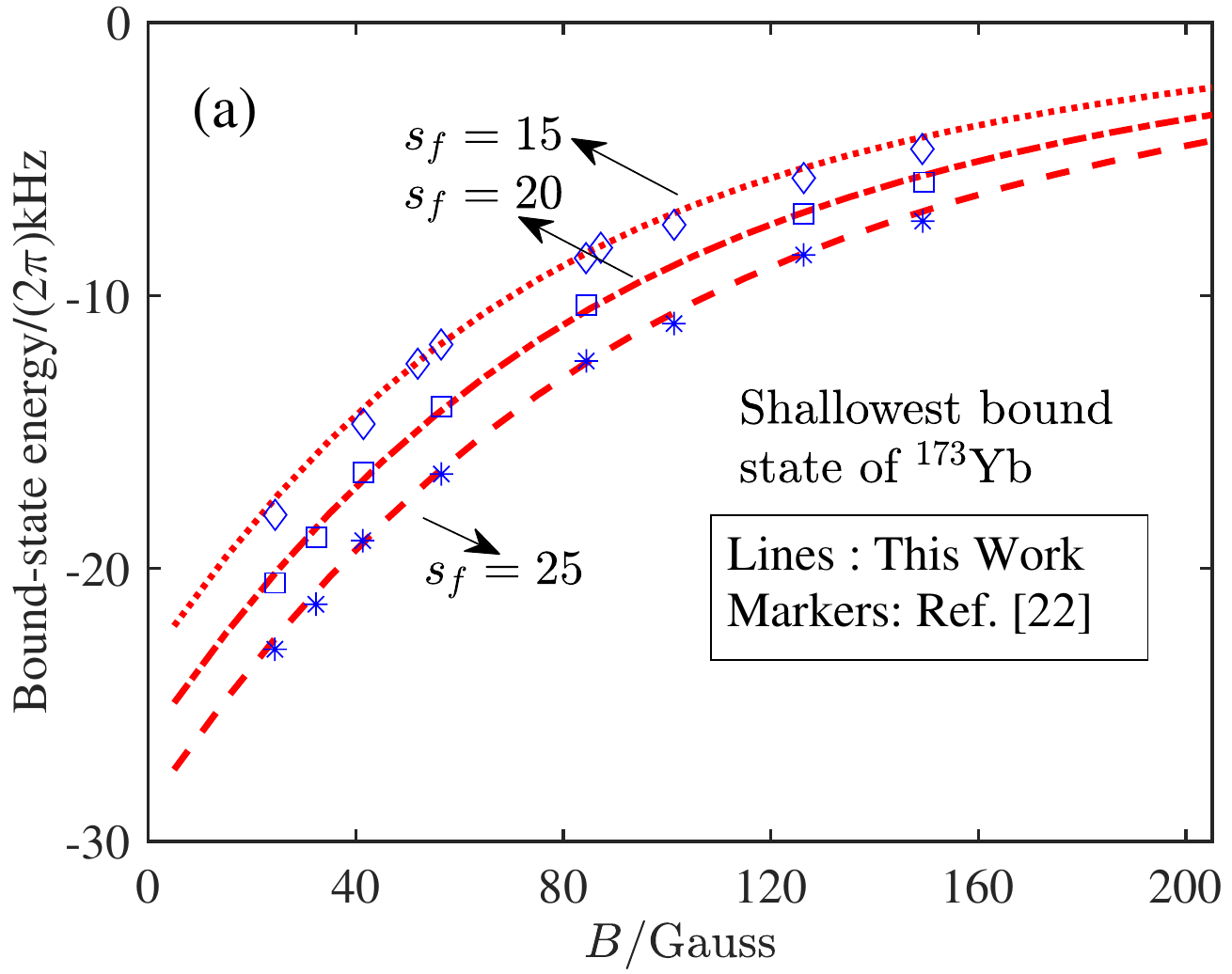} 
\includegraphics[scale=0.4]{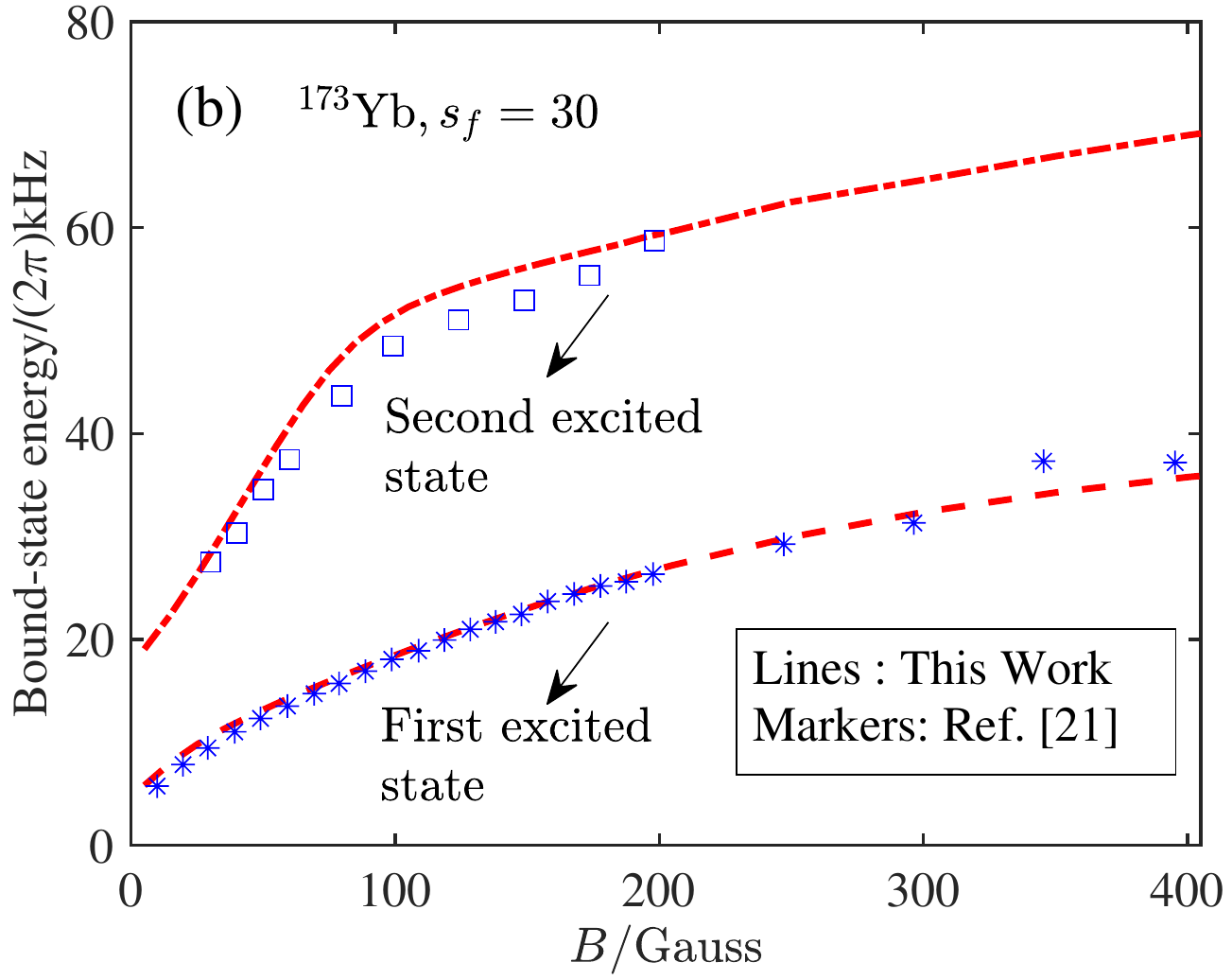}
\includegraphics[scale=0.4]{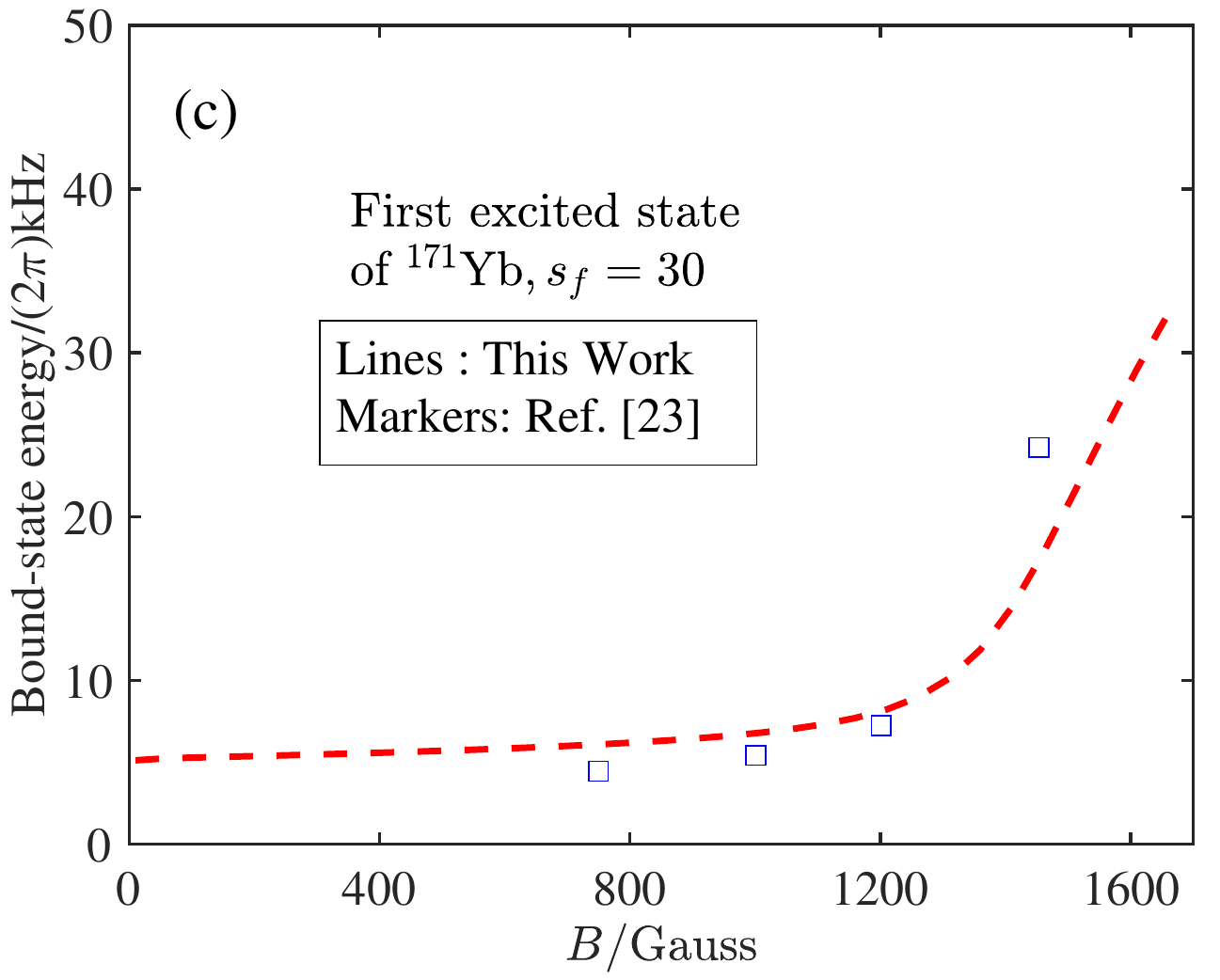}
\caption{(color online) {\bf (a):} The shallowest bound energy spectrum of ultracold $^{173}\rm{Yb}$ atoms with the lattice depth $s_f=15,20,25$. {\bf (b):} The energy spectrum of two excited states
of ultracold $^{173}\rm{Yb}$ atoms with lattice depth $s_f=30$. {\bf (c):} The energy spectrum of one excited state
of ultracold $^{171}\rm{Yb}$ atoms with lattice depth $s_f=30$. 
In the figure we take $E_g^{\rm (c.m.)}=0$, where $E_g^{\rm (c.m.)}=0$ is the two-body ground state energy without the interatomic interactions. Here we show  
 the results given by our calculations with the method in Sec. \ref{sec:sec3} (solid lines) for the best-fitting parameters $a_{+}=2012a_{0}$, $a_{-}=193a_{0}$ ($^{173}{\rm {Yb}}$) and $a_{+}=232a_{0}$, $a_{-}=372a_{0}$ ($^{171}{\rm {Yb}}$), as well as the experimental results given by Refs. \cite{PRL2015,PRX2019} ($^{173}{\rm {Yb}}$) and Ref. \cite{Bettermann} ($^{171}{\rm {Yb}}$) (stars, open circles, open squares and open diamonds). }
\label{fig:Eb}
\end{figure*}

In Fig.~\ref{fig:YbFit} we illustrate the bound-state energies given by our calculation with the above best-fitting parameters (solid lines) and the corresponding experimental results of Refs. \cite{PRL2015,PRX2019,Bettermann}
(open circles and stars). It is clearly shown that they quantitatively agree with each other. To indicate the scattering lengths variability of Yb atoms, we further plot a range of $\pm10\%$ variation of $a_{\pm}$ as the blue shaded areas in Fig.~\ref{fig:YbFit}. 
For the $^{173}{\rm Yb}$ atoms, the energy spectrum is insensitive to the short-range parameters, which is consistent with the observation in Ref.~\cite{PRL2015}. By contrast, the energy spectrum of the $^{171}{\rm Yb}$ atoms is very sensitive to the variations of the scattering lengths.

We further 
use the best-fitted values of $a_{\pm}$ obtained above to
calculate the shallowest bound energies for 
$^{173}{\rm {Yb}}$ atoms
with different lattice depth $s_{f}=15,20,25$, as well as the energies of several excited states of $^{173}{\rm {Yb}}$ atoms
or $^{171}{\rm {Yb}}$ atoms
with $s_{f}=30$. In Fig.~\ref{fig:Eb} we compare our result with the experimental results of Refs.~\cite{PRL2015,PRX2019,Bettermann}.
It is shown that the theoretical (red lines) and experimental results (blue markers) consist very well with each other.

As mentioned before, in Refs. \cite{PRL2015,PRX2019} and Ref. \cite{Bettermann} 
 the values of $a_{\pm}$ were also derived
for $^{173}$Yb and $^{171}$Yb atoms, respectively, via the fitting of the theoretical results with experimental measurements of two-body eigen-energies. Nevertheless, in these calculations the $r$-independent Zeeman Hamiltonian $\hat{h}_S(\delta)$ was ignored in the short-range limit $r\rightarrow 0$, which means the Zeeman Hamiltonian $\hat{h}_S(\delta)$ was omitted in Eqs. (\ref{see}) and (\ref{see2}), or the coefficients $a_{lj}(E,\delta)$ were assumed as $a_{lj}(E,\delta=0)$.
In Tab.~\ref{tab:tabI} we summarize the values of $a_{\pm}$ given by 
these works as well as our above results.
It is shown that our calculation calibrates the values of $a_\pm$ $3\%-12\%$ different from those given by previous works.

\section{Conclusions and outlook}
\label{sec:sec5}

In this work we develop a generic energy-dependent HYP approach for the two-body problem of confined ultracold atoms. 
Instead of directly solving the Schr\"odinger equation of a two-body problem, we encapsulate the interatomic interaction potential to an energy-dependent HYP, which reproduces the wave function of the two-body problem in the short-range region. 
The energy-dependent HYP is characterized by a ``scattering length operator", which self-consistently takes account of the c.m.-relative coupling and the non-commutativity between the internal state Hamiltonian of atoms and the inter-atomic interaction. 
In addition, when the inter-atomic interaction can be approximated as an internal-state independent van der Waals potential beyond a specific range, the scattering length operator can be analytically derived via the multi-channel MQDT approach.
Using our approach, we further calculate the energy spectrum of two alkali-earth (like) atoms confined in a site of a 3D optical lattice.
By fitting the calculated results with the experimentally-measured bound-state energy, we calibrate the values of the zero-energy scattering lengths $a_\pm$ of the short-range inter-atomic interaction at most $12\%$ different from previous works. 
Without introducing extra parameters, our theory unifies several experimental results of the alkali-earth (like) atoms and thereby provides a general framework to deal with related issues.

\begin{acknowledgments}
This work is supported in part by the National Key Research and
Development Program of China Grants No. 2018YFA0306502 (PZ) and No. 2018YFA0307601(RZ),
NSFC Grant No.12174300(RZ), NSAF Grant No. U1930402, as well as the Research Funds of Renmin University of China
under Grant No. 16XNLQ03(PZ).
\end{acknowledgments}

\begin{widetext}

\appendix

\section{MQDT Calculation}

\label{sec:appA}

In this appendix, we show how to derive the parameters $a_{lj}(E,\delta)$ and $A_{lj}(E,\delta)$ introduced in Sec.~\ref{sec:sec2B2} and Sec.~\ref{sec:sec2B4}, respectively, via the MQDT method.

\subsection{Derivation of $a_{lj}(E,\delta)$ of Sec.~\ref{sec:sec2B2}}
\label{appqdt}

 For clarity, here we show the calculation for a specific example (i.e., the system studied in Sec. \ref{sec:sec3}). The generalization of the calculation to other systems is straightforward.
We consider the system with a two-dimensional internal-state space ${\mathscr H}_{S}$ (i.e., $N_S=2$), which has an orthonormal basis $\{|c\rangle_S,|o\rangle_S\}$, and assume the Hamiltonian ${\hat h}_S$ 
and the bare inter-atomic interaction potential ${ U}_{\rm bare}(r)$ are
 given by
\begin{eqnarray}
{\hat h}_S=\delta|c\rangle_S\langle c|,
\end{eqnarray}
and
\begin{eqnarray}
{ U}_{\rm bare}(r)=U_{+}(r)|+\rangle_S\langle+|+U_{-}(r)|-\rangle_S\langle-|,
\end{eqnarray}
respectively. Here $\delta$ is the external-field parameter and the states $|\pm\rangle_S$ are defined as
\begin{eqnarray}
|\pm\rangle_S=\frac{1}{\sqrt{2}}\left(\vert c\rangle_S\mp\vert o\rangle_S\right).\label{pm}
\end{eqnarray}
Moreover, as shown in Sec.~\ref{sec:sec2B2}, we assume that beyond a particular range $b$
both $U_{+}(r)$ and $U_{-}(r)$ can be approximated as van der Waals potentials with the same characteristic length $\beta_{6}$ , i.e.,
\begin{equation}
{U}_{+}\left(r>b\right)\approx{U}_{-}\left(r>b\right)\approx-\frac{\beta_{6}^{4}}{2\mu r^{6}}.\label{fac}
\end{equation}
Namely, we have $U_{+}(r)\neq U_{-}(r)$ only for $r<b$.

Now we use MQDT to solve Eq. (\ref{e3a}), i.e, the equation
\begin{eqnarray}
\left\{-\frac{1}{2\mu}\frac{1}{r^2}\frac{d}{dr}\left(r^2\frac{d}{dr}\right)+{\hat h}_S+{\hat U}_{{\rm bare}}(r)\right\} |\psi(r)\rangle_S=E|\psi(r)\rangle_S,
\label{e3aa}
\end{eqnarray}
 with boundary condition $|\psi(r=0)\rangle_S=0$.
 This has already been done in our previous work \cite{Zhang2017}. Thus, here we just briefly show the principle of the MQDT method and the main results. More details of the MQDT calculations are shown in Ref. \cite{Zhang2017}.

 We first consider the case with $\delta=0$, where the $|+\rangle_S$ and $|-\rangle_S$ components of Eq. (\ref{e3aa}) are decoupled with each other. Due to the above fact (\ref{fac}), the solution of Eq. (\ref{e3aa}) satisfies (up to a global constant)
 \begin{eqnarray}
 _S\langle \pm|\psi(r)\rangle_S & = & \frac{1}{r}\left[f_{E}^{0}(r)-K_{\pm}^{0}g_{E}^{0}(r)\right],
 \hspace{0.8cm}
 ({\rm for}\ r>b).\label{psip}
 \end{eqnarray}
 Here $f_{E}^{0}(r)$ and $g_{E}^{0}(r)$ are two linearly-independent special solutions of the single-component
 radial Schr\"odinger equation with van der Waals potential:
 \begin{eqnarray}
 \left\{-\frac{1}{2\mu}\frac{d^2}{dr^2}-\frac{1}{2\mu}\frac{\beta_{6}^{4}}{r^{6}} \right\} y(r)=Ey(r),
 \label{rse}
 \end{eqnarray}
 with eigen-value $E$, which were derived by Bo Gao in \cite{Gao1998,Gao1998QDT}. Moreover, in Eq. (\ref{psip}) the parameters $K_{+}^{0}$ and $K_{-}^{0}$ are determined by the details of the potentials $U_+(r)$ and $U_-(r)$ in the region $r<b$, respectively.
 Here we emphasisze that, the functions $f_{E}^{0}(r)$ and $g_{E}^{0}(r)$ are chosen to satisfy energy-independent normalization conditions in the limit $r\rightarrow 0$ \cite{Gao1998QDT,Gao2005}. 
In the low-energy cases with $E$ being much less than the van der Waals energy $1/(2\mu\beta_6^2)$, both the functions $f_{E}^{0}(r)$ and $g_{E}^{0}(r)$ and the function $_S\langle \pm|\psi(r)\rangle_S$ (up to a global factor) are almost independent of $E$ in the region $r\approx b$. As a result, the parameters $K_{\pm}^{0}$ are also almost independent of $E$. It was proved that these two parameters are related to the zero-energy scattering lengths $a_{\pm}$ with respect to the potentials $U_{\pm}(r)$ via
 \begin{equation}
 K_{\pm}^{0}=\frac{2\pi\beta_{6}}{2\pi\beta_{6}-a_{\pm}\Gamma\left(1/4\right)^{2}},
 \end{equation}
 with $\Gamma\left(z\right)$ being the Gamma function \cite{Gao1998QDT}.

 Now we consider the cases with $\delta\neq 0$ and $|\delta|\ll 1/(2\mu\beta_6^2)$. Also due to the fact Eq. (\ref{fac}), the solution of Eq. (\ref{e3aa}) satisfies
 \begin{eqnarray}
 |\psi(r)\rangle_S=\frac{1}{r} \left[A_{f}f_{E}^{0}\left(r\right)+A_{g}g_{E}^{0}\left(r\right)\right]\vert o\rangle_S+\frac{1}{r} \left[B_{f}f_{E-\delta}^{0}\left(r\right)+B_{g}g_{E-\delta}^{0}\left(r\right)\right]\vert c\rangle_S,\hspace{0.6cm}({\rm for}\ r>b),\label{ss}
 \end{eqnarray}
 with $A_{f,g}$ and $B_{f,g}$ being $r$-independent coefficients.
 On the other hand, in our low-energy case $f_{E}^{0}(r)$ and $g_{E}^{0}(r)$ are almost independent of $E$ for $r\approx b$. Thus, we have
 \begin{eqnarray}
 |\psi(r)\rangle_S\approx\frac{1}{r} \left[A_{f}f_{E}^{0}\left(r\right)+A_{g}g_{E}^{0}\left(r\right)\right]\vert o\rangle_S+\frac{1}{r} \left[B_{f}f_{E}^{0}\left(r\right)+B_{g}g_{E}^{0}\left(r\right)\right]\vert c\rangle_S,\hspace{0.6cm}({\rm for}\ r\approx b ).\label{psip3}
 \end{eqnarray}
 Furthermore, as shown in the main text below Eq.~(\ref{vd}), the van der Waals characteristic length $\beta_6$ satisfies the low-energy condition $|k_j|^2\ll1/\beta_6^2$ ($j=1,...,N_S$), which can be expressed as $(|E|, |\delta|)\ll 1/(2\mu\beta_6^2)$ for our current system. This condition yields that the behavior of $|\psi(r)\rangle_S$ in the region $r\approx b$ is almost independent of $\delta$. Using this fact and the expression Eq. (\ref{psip}) for the wave function for $\delta=0$, we find that even for finite $\delta$ we still have
 \begin{eqnarray}
 _S\langle \pm|\psi(r)\rangle_S & \approx & \frac{1}{r}\left[f_{E}^{0}(r)-K_{\pm}^{0}g_{E}^{0}(r)\right],
 \hspace{0.8cm}
 ({\rm for}\ r\approx b).\label{psip2}
 \end{eqnarray}
 The above results (\ref{psip3}) and (\ref{psip2}) are key points of the MQDT approach. Combining these two equations we can obtain the algebraic equations which must be satisfied by the coefficients $A_{f,g}$ and $B_{f,g}$. With further direct calculations based on these algebraic equations and Eqs. (\ref{pm}), ( \ref{ss}, \ref{psip2}) and (\ref{psip2}), we further find the behaviors of two linearly-independent special solutions of Eq. (\ref{e3aa}) in the region $r>b$:
 \begin{eqnarray}
 \vert\psi^{\left(\alpha\right)}\left(r>b\right)\rangle_S
 &=& \frac{1}{r}\left\{ \left[f_{E}^{0}\left(r\right)-K_{oo}^{0}g_{E}^{0}\left(r\right)\right]\vert o\rangle_S-K_{co}^{0}g_{E-\delta}^{0}\left(r\right)\vert c\rangle_S\right\} ,\\
 \vert\psi^{\left(\beta\right)}\left(r>b\right)\rangle_S
 &=& \frac{1}{r}\left\{ -K_{oc}^{0}g_{E}^{0}\left(r\right)\vert o\rangle_S+\left[f_{E-\delta}^{0}\left(r\right)-K_{cc}^{0}g_{E-\delta}^{0}\left(r\right)\right]\vert c\rangle_S\right\} ,\label{psibeta}
 \end{eqnarray}
 with the coefficients $K_{ij}^{0}$ ($i,j=o,c$) being defined as
 \begin{eqnarray}
 K_{oo}^{0} = K_{cc}^{0}=\frac{K_{+}^{0}+K_{-}^{0}}{2};\hspace{0.8cm}
 K_{co}^{0} = K_{oc}^{0}=\frac{K_{-}^{0}-K_{+}^{0}}{2}.
 \end{eqnarray}

 Above we have solved Eq. (\ref{e3aa}) with MQDT. Now we use these results to derive the parameters $a_{lj}(E,\delta)$. To this end, we should use $\vert\psi^{\left(\alpha,\beta\right)}\left(r\right)\rangle_S$ to compose another two special solutions
 of Eq. (\ref{e3aa}), which are introduced in Eq. (\ref{psi12}). Explicitly, we require to solve the equations
 \begin{eqnarray}
C_\alpha\vert\psi^{\left(\alpha\right)}\left(r\gtrsim \beta_6\right)\rangle_S+C_\beta\vert\psi^{\left(\beta\right)}\left(r\gtrsim \beta_6\right)\rangle_S
 \propto\frac{1}{r}
 \bigg\{
 \frac{1}{k_o}\sin(k_o r)|o\rangle_S -
 \Big[
 a_{co}(E,\delta)\cos(k_{c} r)|c\rangle_S+a_{oo}(E,\delta)\cos(k_{o} r)|o\rangle_S\Big]\bigg\};\nonumber\\
 \label{f1}
 D_\alpha\vert\psi^{\left(\alpha\right)}\left(r\gtrsim \beta_6\right)\rangle_S+D_\beta\vert\psi^{\left(\beta\right)}\left(r\gtrsim \beta_6\right)\rangle_S
 \propto\frac{1}{r}
 \bigg\{
 \frac{1}{k_c}\sin(k_c r)|c\rangle_S -
 \Big[
 a_{cc}(E,\delta)\cos(k_{c} r)|c\rangle_S+a_{oc}(E,\delta)\cos(k_{o} r)|o\rangle_S\Big]\bigg\},\nonumber\\
 \label{f2}
 \end{eqnarray}
 where $k_c=\sqrt{2\mu(E-\delta)}$ and $k_o=\sqrt{2\mu E}$, with $C_{\alpha,\beta}$, $D_{\alpha,\beta}$ and $a_{lj}(E,\delta)$ ($l,j=o,c$) being the unknowns. This equation can be solved with the asymptotic behavior of the functions $f_\epsilon^0(r)$ and $g_\epsilon^0(r)$ ($\epsilon=E,E-\delta$) in the region $r\gtrsim \beta_6$ where the van der Waals interaction can be ignored. These behaviors were provided by Gao in Refs. \cite{Gao2005,Gao2007}. According to these references, we can introduce another two solutions
 $\{f_\epsilon^c(r),g_\epsilon^c(r)\}$ of Eq. (\ref{rse}) which are related to $\{f_\epsilon^0(r),g_\epsilon^0(r)\}$ via
 \begin{equation}
 \left(\begin{array}{c}
 f_{\epsilon}^{0}\left(r\right)\\
 g_{\epsilon}^{0}\left(r\right)
 \end{array}\right)=\sqrt{2}\left(\begin{array}{cc}
 \cos\left(\frac{\pi}{8}\right) & -\sin\left(\frac{\pi}{8}\right)\\
 -\sin\left(\frac{\pi}{8}\right) & -\cos\left(\frac{\pi}{8}\right)
 \end{array}\right)\left(\begin{array}{c}
 f_{\epsilon}^{c}\left(r\right)\\
 g_{\epsilon}^{c}\left(r\right)
 \end{array}\right).
 \label{tr1}
 \end{equation}
 Furthermore, the behaviors of $\{f_\epsilon^c(r),g_\epsilon^c(r)\}$ for $r\gtrsim\beta_6$ are \cite{Gao2007}:
 \begin{equation}
 \left(\begin{array}{c}
 f_\epsilon^{c}\left(r\gtrsim\beta_{6}\right)\\
 g_\epsilon^{c}\left(r\gtrsim\beta_{6}\right)
\end{array}\right)\approx\sqrt{\frac{2}{\pi k_\epsilon\beta_{6}}}\left(\begin{array}{cc}
 Z_{ff}^{c\left(6\right)} & Z_{fg}^{c\left(6\right)}\\
 Z_{gf}^{c\left(6\right)} & Z_{gg}^{c\left(6\right)}
 \end{array}\right)\left(\begin{array}{c}
 \sin\left( k_\epsilon r\right)\\
 -\cos\left( k_\epsilon r\right)
 \end{array}\right),\hspace{0.5cm}({\rm for}\ \epsilon>0),\label{c1}
 \end{equation}
 and
 \begin{equation}
 \left(\begin{array}{c}
 f_{\epsilon}^{c}\left(r\gtrsim\beta_{6}\right)\\
 g_{\epsilon}^{c}\left(r\gtrsim\beta_{6}\right)
\end{array}\right)\approx\sqrt{\frac{2}{\pi k_\epsilon \beta_{6}}}\left(\begin{array}{cc}
 \frac{W_{f-}^{c\left(6\right)}+2W_{f+}^{c\left(6\right)}}{2} & \frac{W_{f-}^{c\left(6\right)}-2W_{f+}^{c\left(6\right)}}{2}\\
 \frac{W_{g-}^{c\left(6\right)}+2W_{g+}^{c\left(6\right)}}{2} & \frac{W_{g-}^{c\left(6\right)}-2W_{g+}^{c\left(6\right)}}{2}
 \end{array}\right)\left(\begin{array}{c}
 \sinh(k_\epsilon r)\\
 \cosh(k_\epsilon r)
 \end{array}\right),\hspace{0.5cm}({\rm for}\ \epsilon<0),\label{c2}
 \end{equation}
 where $k_\epsilon=\sqrt{2\mu\vert\epsilon\vert}$, and $Z_{ij}^{c\left(6\right)}\left(i,j=f,g\right)$ and $W_{ij}^{c\left(6\right)}\left(i=f,g;j=\pm\right)$
 are also functions of $\epsilon$ and are listed in \cite{Gao2007,Gao2000}. 
 Substituting Eqs. (\ref{c1}) and (\ref{c2}) into Eq. (\ref{tr1}), we can derive the behaviors of the functions $f_\epsilon^0(r)$ and $g_\epsilon^0(r)$ ($\epsilon=E,E-\delta$) in the region with $r\gtrsim \beta_6$. Using these behaviors, we can directly solve Eq. (\ref{f2}) and obtain
{
\begin{eqnarray}
 a_{oo}(E,\delta)&=&\frac{1}{\left|k_{o}\right|}\cdot\frac{\left(P_{f,E-\delta}-K_{cc}^{0}P_{g,E-\delta}\right)\left(Q_{f,E}-K_{oo}^{0}Q_{g,E}\right)-K_{co}^{0}K_{oc}^{0}P_{g,E-\delta}Q_{g,E}}{K_{co}^{0}K_{oc}^{0}P_{g,E-\delta}P_{g,E}-\left(P_{f,E}-K_{oo}^{0}P_{g,E}\right)\left(P_{f,E-\delta}-K_{cc}^{0}P_{g,E-\delta}\right)}, \\
 \nonumber\\
 a_{oc}(E,\delta)&=&-\frac{1}{\left|k_{c}\right|}\frac{\left(P_{f,E}-K_{oo}^{0}P_{g,E}\right)K_{oc}^{0}Q_{g,E}-K_{oc}^{0}P_{g,E}\left(Q_{f,E}-K_{oo}^{0}Q_{g,E}\right)}{K_{co}^{0}K_{oc}^{0}P_{g,E-\delta}P_{g,E}-\left(P_{f,E}-K_{oo}^{0}P_{g,E}\right)\left(P_{f,E-\delta}-K_{cc}^{0}P_{g,E-\delta}\right)}, \\
 \nonumber\\
 a_{co}(E,\delta)&=&-\frac{1}{\left|k_{o}\right|}\frac{\left(P_{f,E-\delta}-K_{cc}^{0}P_{g,E-\delta}\right)K_{co}^{0}Q_{g,E-\delta}-K_{co}^{0}P_{g,E-\delta}\left(Q_{f,E-\delta}-K_{cc}^{0}Q_{g,E-\delta}\right)}{K_{co}^{0}K_{oc}^{0}P_{g,E-\delta}P_{g,E}-\left(P_{f,E}-K_{oo}^{0}P_{g,E}\right)\left(P_{f,E-\delta}-K_{cc}^{0}P_{g,E-\delta}\right)},\\
 \nonumber\\
a_{cc}(E,\delta)&=&\frac{1}{\left|k_{c}\right|}\frac{\left(Q_{f,E}-K_{oo}^{0}Q_{g,E}\right)\left(Q_{f,E-\delta}-K_{cc}^{0}Q_{g,E-\delta}\right)-\left(-K_{oc}^{0}P_{g,E}\right)\left(-K_{co}^{0}Q_{g,E-\delta}\right)}{K_{co}^{0}K_{oc}^{0}P_{g,E-\delta}P_{g,E}-\left(P_{f,E}-K_{oo}^{0}P_{g,E}\right)\left(P_{f,E-\delta}-K_{cc}^{0}P_{g,E-\delta}\right)},
\end{eqnarray}
}
where $P_{f,\epsilon},Q_{f,\epsilon},P_{g,\epsilon},Q_{g,\epsilon}$ are defined as
\begin{align}
 & P_{f,\epsilon}=\sqrt{\frac{4}{\pi k_\epsilon \beta_6}}\left( \cos\left(\frac{\pi}{8}\right)Z_{ff}^{c\left(6\right)}-\sin\left(\frac{\pi}{8}\right)Z_{gf}^{c\left(6\right)}\right),\label{eq:Pf1}\\
 & Q_{f,\epsilon}=\sqrt{\frac{4}{\pi k_\epsilon \beta_6}}\left(-\cos\left(\frac{\pi}{8}\right)Z_{fg}^{c\left(6\right)}+\sin\left(\frac{\pi}{8}\right)Z_{gg}^{c\left(6\right)}\right),\label{eq:Qf1}\\
 & P_{g,\epsilon}=\sqrt{\frac{4}{\pi k_\epsilon \beta_6}}\left(-\sin\left(\frac{\pi}{8}\right)Z_{ff}^{c\left(6\right)}-\cos\left(\frac{\pi}{8}\right)Z_{gf}^{c\left(6\right)}\right),\label{eq:Pg1}\\
 & Q_{g,\epsilon}=\sqrt{\frac{4}{\pi k_\epsilon \beta_6}}\left(\sin\left(\frac{\pi}{8}\right)Z_{fg}^{c\left(6\right)}+\cos\left(\frac{\pi}{8}\right)Z_{gg}^{c\left(6\right)}\right)\label{eq:Qg1}
\end{align}
for $\epsilon>0$, and
\begin{align}
 & P_{f,\epsilon}=\sqrt{\frac{4}{\pi k_\epsilon \beta_6}}\left(\cos\left(\frac{\pi}{8}\right)\frac{W_{f-}^{c\left(6\right)}+2W_{f+}^{c\left(6\right)}}{2}-\sin\left(\frac{\pi}{8}\right)\frac{W_{g-}^{c\left(6\right)}+2W_{g+}^{c\left(6\right)}}{2}\right),\label{eq:Pf2}\\
 & Q_{f,\epsilon}=\sqrt{\frac{4}{\pi k_\epsilon \beta_6}}\left(\cos\left(\frac{\pi}{8}\right)\frac{W_{f-}^{c\left(6\right)}-2W_{f+}^{c\left(6\right)}}{2}-\sin\left(\frac{\pi}{8}\right)\frac{W_{g-}^{c\left(6\right)}-2W_{g+}^{c\left(6\right)}}{2}\right),\label{eq:Qf2}\\
 & P_{g,\epsilon}=\sqrt{\frac{4}{\pi k_\epsilon \beta_6}}\left(-\sin\left(\frac{\pi}{8}\right)\frac{W_{f-}^{c\left(6\right)}+2W_{f+}^{c\left(6\right)}}{2}-\cos\left(\frac{\pi}{8}\right)\frac{W_{g-}^{c\left(6\right)}+2W_{g+}^{c\left(6\right)}}{2}\right),\label{eq:Pg2}\\
 & Q_{g,\epsilon}=\sqrt{\frac{4}{\pi k_\epsilon \beta_6}}\left(-\sin\left(\frac{\pi}{8}\right)\frac{W_{f-}^{c\left(6\right)}-2W_{f+}^{c\left(6\right)}}{2}-\cos\left(\frac{\pi}{8}\right)\frac{W_{g-}^{c\left(6\right)}-2W_{g+}^{c\left(6\right)}}{2}\right)\label{eq:Qg2}
\end{align}
for $\epsilon<0$.

\subsection{Derivation of $A_{lj}(E,\delta)$ of Sec.~\ref{sec:sec2B4}}
\label{appA2}

By straightforwardly generalizing the calculation in Appendix~\ref{appqdt}, one can also derive the coefficients $A_{lj}(E,\delta)$ of Sec.~\ref{sec:sec2B4} via MQDT. Here we just show the main steps of this approach.
For convenience, we assume the Hilbert spaces ${\mathscr H}_{R}$ and ${\mathscr H}_{S}$ for the c.m. motion and internal state are spanned by the basis $\{|\nu\rangle_S|\nu=1,...,N_S\}$ and $\{|\eta\rangle_R|\eta=1,...,N_R\}$, respectively. It is clear that we have $N_{RS}=N_RN_S$.
We further assume the inter-atomic interaction potential $\hat{U}_{\rm bare}(r)$ are diagonal in the basis $\{|\nu\rangle_S|\nu=1,...,N_S\}$ and can be approximated as the van der Waals potential beyond the range $b$, i.e.
\begin{equation}
\hat{U}_{{\rm bare}}\left(r\right)=\begin{dcases*}
\sum_{\nu=1}^{N_{S}}U_{\nu}\left(r\right)\vert\nu\rangle_S\langle\nu\vert & $r<b$\\
-\frac{\beta_{6}^{4}}{2\mu r^{6}} & $r\ge b$
\end{dcases*}.
\end{equation}

Using MQDT we can obtain $N_{RS}$ special solutions of the stationary 
Schrödinger equation of the Hamiltonian $[-{\nabla_{{\bf r}}^{2}}/{(2\mu)}+{\hat U}_{{\rm bare}}(r)+\hat{h}_S]|\psi(r)\rangle_{RS}=E|\psi(r)\rangle_{RS}$, which can be denoted as $\vert\psi_{\xi}(r)\rangle_{RS}$ ($\xi=1,...,N_{RS}$) and satisfy 
\begin{equation}
 \vert\psi_\xi\left(r>b\right)\rangle_{RS}
 = \frac{1}{r}\sum_{j=1}^{N_{RS}} \left[\delta_{\xi j}f_{E_j}^{0}\left(r\right)+A_{g}^{(\xi j)}g_{E_j}^{0}\left(r\right)\right] \vert \lambda_j\rangle_{RS},\ \ (\xi=1,...,N_{RS}),\label{a32}
 \end{equation}
where $E_j=E-\lambda_j$, $\{ \lambda_j, \vert \lambda_j\rangle_{RS} \}$ is defined in Eq. (\ref{nn}), 
$\delta_{\xi j}$ is the Kronecker symbol, 
and the coefficients $A_{g}^{(\xi j)}$ $(\xi,j=1,...,N_{RS})$ are the solutions of the $N_{RS}^2$ equations
\begin{eqnarray}
&&\sum_{j=1}^{N_{RS}} \Big\{A_{g}^{(\xi j)} \big[_R\langle\eta|_S\langle\nu|\lambda_j\rangle_{RS}\big]\Big\}=-K_{\nu}^{0}\big[_R\langle\eta|_S\langle\nu|\lambda_\xi\rangle_{RS}\big],\nonumber\\
\nonumber\\
&&({\rm for}\ \xi=1,...,N_{RS};\ \nu=1,...,N_S;\ \eta=1,...,N_R).
\label{eq:MCE}
\end{eqnarray}
Here $K_{\nu}^{0}$ ($\nu=1,...,N_S$) is given by 
 \begin{equation}
 K_{\nu}^{0}=\frac{2\pi\beta_{6}}{2\pi\beta_{6}-a_{\nu}\Gamma\left(1/4\right)^{2}};\ \ (\nu=1,\cdots,N_{S}),
 \end{equation}
 where $a_\nu$ is the zero-energy scattering lengths with respect to the potential curve $_S\langle \nu |{\hat U}_{\rm bare}(r)|\nu\rangle_S$. 
 
As in Sec.~\ref{sec:sec2B2}, using the special solutions $\vert\psi_{\xi}(r)\rangle_{RS}$ ($\xi=1,...,N_{RS}$) 
obtained above
 we can derive the coefficients $A_{lj}(E,\delta)$ of Sec.~\ref{sec:sec2B4}
via solving the algebraic equations 
\begin{equation}
\sum_{l=1}^{N_{RS}} D_{l}^{(j)} \vert\psi_l\left(r\gtrsim\beta_6\right)\rangle_{RS} \propto \frac{1}{r}
\left\{
\frac{1}{p_j}\sin(p_j r)|\lambda_j\rangle_{RS} -
\sum_{l=1}^{N_{RS}}
A_{lj}(E,\delta)
\cos(p_{l} r)|\lambda_{l}\rangle_{RS}\right\};\ \ (j=1,\cdots,N_{RS}),
\end{equation}
where 
the long-range wave functions $\vert\psi^{\left(l\right)}\left(r\gtrsim\beta_6\right)\rangle_{RS}$ ($l=1,...,N_{RS}$) are given by substituting Eqs.~(\ref{c1}) and (\ref{c2}) into Eq.~(\ref{a32}), 
$p_l=\sqrt{2\mu(E-\lambda_l)}$ ($l=1,...,N_{RS}$), 
and $D_l^{(j)},A_{lj}(E,\delta)$ are unknowns.

\section{Solution of Eq. (\ref{err2})}
\label{sec:appB}

In this appendix, we solve Eq. (\ref{err2}), i.e., the equation
$
 {\hat H}^\prime_{\rm eff}(E_g^{(0)})|\Psi(r)\rangle_{RS}=E_g^{(0)}\,|\Psi(r)\rangle_{RS},
$
under the boundary condition $|\Psi(r\rightarrow\infty)\rangle_{RS}=0$,
 and derive the zero-th order result $E_g^{(0)}$ for the ground-state energy of the two confined alkaline-earth (like) atoms.

As shown Sec.~\ref{sec:sec3C}, the term $\hat{ V}_1^{\rm (c)}$ is neglected in $ {\hat H}^\prime_{\rm eff}$, and thus
the total confinement potential is approximated as
\begin{eqnarray}
{ V}^{\rm (c)}({\hat {\bf R}}, {\bf r})\approx
\frac{1}{2}\mu\omega^{2}r^{2}+V_{0R}^{\rm (c)}({\hat {\bf R}}),
\end{eqnarray}
which does not include the coupling between the relative and c.m. motion. Therefore, to solve Eq. (\ref{err2}) we can separate these two degrees of freedoms. Explicitly, with straightforward calculation, we find that the solution $\{E_g^{(0)},|\Psi(r)\rangle_{RS}\}$ of Eq. (\ref{err}) can be expressed as
\begin{eqnarray}
|\Psi(r)\rangle_{RS}&=&|\psi_{\rm rel}(r)\rangle_S|{\cal E}_g\rangle_R;\\
E_g^{(0)}&=&E_{{\rm r}g}+{\cal E}_g,\hspace{0.5cm}\label{ebb}
\end{eqnarray}
Here $|{\cal E}_g\rangle_R$ and ${\cal E}_g$ are the ground-state and the corresponding eigen-energy of the c.m. Hamiltonian $\hat{ H}_R$ defined in Eq. (\ref{hr}). In addition, $\{E_{{\rm r}g},|\psi_{\rm rel}(r)\rangle_S\}$ are the ground-state solutions of the Schr\"odinger equation for the relative motion:
\begin{eqnarray}
 \left[\left(-\frac{1}{2\mu}\nabla_{{\bf r}}^{2}+\frac{1}{2}\mu\omega^{2}r^{2}\right)+\delta\,\vert c\rangle_S\langle c\vert+
 \sum_{l,j=o,c} |l\rangle_S\langle j| a_{lj}(E_{{\rm r}g},\delta)\frac{2\pi}{\mu}\delta({\bf r})\frac{\partial}{\partial r}(r\cdot)
 \right]|\psi_{\rm rel}(r)\rangle_S=E_{{\rm r}g}|\psi_{\rm rel}(r)\rangle_S,\label{eraa}
\end{eqnarray}
in the $s$-wave manifold with boundary condition $\vert\psi_{\rm rel}(r\rightarrow\infty)\rangle_S=0$, with the coefficients $a_{lj}(E_{{\rm r}g},\delta)$ ($l,j=o,c$) derived in Sec.~\ref{sec:sec3B}.

We can solve Eq. (\ref{eraa})
by straightforwardly generalizing the seminal work of T. Bush \cite{Busch1998}. To this end, we first re-express Eq. (\ref{eraa}) in the region with $r>0$ as
\begin{eqnarray}
\left(-\frac{1}{2\mu}\frac{d^{2}}{dr^{2}}+\frac{1}{2}\mu\omega^{2}r^{2}+\Delta_{j}\right)\big[r\cdot_S\!\langle j|\psi_{\rm rel}(r)\rangle_S\big] & = & E_{{\rm r}g}\big[r\cdot_S\!\langle j|\psi_{\rm rel}(r)\rangle_S\big]
\ \ \ \ \ \ ({\rm for}\ j=o,c),\label{eq:ode}
\end{eqnarray}
with $\Delta_{o}=0$ and $\Delta_{c}=\delta$. Combining this fact
and the binding condition $_S\langle j|\chi(r\rightarrow\infty)\rangle_S=0$,
we find that
\begin{equation}
|\Psi\left(r\right)\rangle_S=\frac{1}{r}\left[c_{1}D_{\nu_{o}}\left(\frac{\sqrt{2}r}{l_{{\rm ho}}}\right)\vert o\rangle_S+c_{2}D_{\nu_{c}}\left(\frac{\sqrt{2}r}{l_{{\rm ho}}}\right)\vert c\rangle_S\right],\label{sol}
\end{equation}
where the characteristic length $l_{{\rm ho}}=\sqrt{{1}/({\mu\omega})}$, $c_{1},c_{2}$ are $r$-independent constants,
$D_{\nu}(z)$ is the parabolic cylinder function and the parameters
$\nu_{o,c}$ are defined as
\begin{eqnarray}
\nu_{o}=\frac{E_{{\rm r}g}}{\omega}-\frac{1}{2};\ \ \nu_{c}=\frac{E_{{\rm r}g}-\delta}{\omega}-\frac{1}{2}.\label{eq:nuc}
\end{eqnarray}
Furthermore, the HYP in Eq. (\ref{eraa}) is mathematically equivalent to the
 two channel Bethe-Perierls boundary condition
\begin{eqnarray}
|\psi_{\rm rel}^{(-1)}\rangle_S & = & -\left[\sum_{i,j=o,c}a_{ij}(E_{{\rm r}g},\delta)\vert i\rangle_S\langle j\vert\right]|\psi_{\rm rel}^{(0)}\rangle_S,\label{bp2}
\end{eqnarray}
with the $r$-independent spin states $|\psi_{\rm rel}^{(-1,0)}\rangle_{S}$ being
the terms in the small-$r$ expansion of $|\psi_{\rm rel}\left(r\right)\rangle_{RS}$:
\begin{eqnarray}
\lim_{r\rightarrow0^{+}}|\psi_{\rm rel}(r)\rangle_{S}=\frac{1}{r}|\psi_{\rm rel}^{(-1)}\rangle_S+|\psi_{\rm rel}^{(0)}\rangle_S+{\cal O}(r).\label{exp2}
\end{eqnarray}
Substituting Eq.~(\ref{sol}) into Eqs.~(\ref{exp2}, \ref{bp2})
and using the fact
\begin{equation}
\lim_{z\rightarrow0^{+}}D_{\nu}\left(z\right)=\frac{\sqrt{2^{\nu}\pi}}{\Gamma(\frac{1}{2}-\frac{\nu}{2})}-\frac{\sqrt{2^{\nu+1}\pi}}{\Gamma(-\frac{\nu}{2})}z+{\cal O}(z^{2}),
\end{equation}
we find that the condition (\ref{bp2}) yields a linear equation for
the coefficients $c_{1,2}$ in the expression (\ref{sol}) of $|\psi_{\rm rel}(r)\rangle$:
\begin{eqnarray}
\mathbb{J}(E_{{\rm r}g})\left(\begin{array}{l}
c_{1}\\
c_{2}
\end{array}\right)=0,\label{le}
\end{eqnarray}
where the $2\times2$ matrix $\mathbb{J}(E_{{\rm r}g})$ is defined as
\begin{eqnarray}
\mathbb{J}(E_{{\rm r}g})=\frac{1}{l_{{\rm ho}}}\left(\begin{array}{cc}
l_{{\rm ho}}\Gamma_{1}^{o}+a_{oo}\Gamma_{2}^{o}, & a_{oc}\Gamma_{2}^{c}\\
a_{co}\Gamma_{2}^{o}, & l_{{\rm ho}}\Gamma_{1}^{c}+a_{cc}\Gamma_{2}^{c}
\end{array}\right)
\end{eqnarray}
with symbol conventions
\begin{equation}
\Gamma_{1}^{j}=\frac{\sqrt{2^{\nu_{j}}\pi}}{\Gamma(\frac{1}{2}-\frac{\nu_{j}}{2})};\ \ \Gamma_{2}^{j}=-\frac{2\sqrt{2^{\nu_{j}}\pi}}{\Gamma(-\frac{\nu_{j}}{2})},\ (\ j=o,c),
\end{equation}
which yields that the energy $E_{{\rm r}\alpha}$ ($\alpha=1,2,...$) satisfy the algebraic equation
\begin{equation}
\det\left[\mathbb{J}(E_{{\rm r}\alpha})\right]=0.\label{eq:JE}
\end{equation}
Thus, by solving Eq. (\ref{eq:JE}) we can obtain all the eigen-energy $E_{{\rm r}\alpha}$ of the relative motion. Substituting this result and the c.m. ground energy $\epsilon_g$ obtained in Sec.~\ref{sec:sec3B} into Eq. (\ref{ebb}), we obtain the zero-th order two-atom ground-state energy $E_g^{(0)}=E_{{\rm r}g}+{\cal E}_g$ with $E_{{\rm r}g}$ being the minimum of $E_{{\rm r}\alpha}$.

\section{Diagonalization of $\hat{H}_{\rm eff}(E_g^{(0)})$ }
\label{sec:appC}

In this appendix we show our method for the diagonalization of the Hamiltonian $\hat{H}_{\rm eff}(E_g^{(0)})$ of Sec. \ref{sec:sec3C}. As shown in the main text, in our calculation the function $\hat{H}_{\rm eff}(E_b)$ is defined in Eq. (\ref{heff}), and the value of the argument $E_g^{(0)}$ are already derived in the previous calculations. Thus, our purpose is to diagonalize a totally determined Hamiltonian $\hat{H}_{\rm eff}(E_g^{(0)})$.

The key point is that $\hat{H}_{\rm eff}(E_g^{(0)})$ includes a HYP term ${\hat a}_{\rm eff}(E_g^{(0)})\frac{2\pi}{\mu}\delta({\bf r})\frac{\partial}{\partial r}(r\cdot)$ with determined parameter $E_g^{(0)}$. Due to this term, the eigen-state $|\Phi(r)\rangle_{RS}$ of the $\hat{H}_{\rm eff}(E_g^{(0)})$ must satisfy the corresponding Bethe Peierls boundary condition (BPC), i.e., in the short-range limit $r\rightarrow 0$, the state $|\Phi(r)\rangle_{RS}$ can be expressed as
\begin{eqnarray}
\lim_{r\rightarrow 0}|\Phi(r)\rangle_{RS}=\left[\frac{{\hat a}_{\rm eff}(E_g^{(0)})}r-1\right]|\chi\rangle_{RS}+{\cal O}(r),\label{BPC}
\end{eqnarray}
with $|\chi\rangle_{RS}$ being a $r$-independent state. Therefore, we should first find a complete orthogonal basis $\{|\phi_{\lambda}(r)\rangle_{RS}|\lambda=0,1,2,...\}$, in which all states satisfy the BPC (i.e. Eq.~(\ref{BPC})), and then 
express $\hat{H}_{\rm eff}(E_g^{(0)})$ as a matrix in this basis, and numerically diagonalize that matrix.

In our calculation we use the eigen-states of $\hat{H}_{\rm eff}^\prime(E_g^{(0)})$ as 
basis $\{|\phi_{\lambda}(r)\rangle_{RS}|\lambda=0,1,2,...\}$, with $\hat{H}_{\rm eff}^\prime$ being defined in Eq. (\ref{err2a}). Explicitly, we first solve the equation
\begin{eqnarray}
\hat{H}_{\rm eff}^\prime(E_g^{(0)})|\phi_{\lambda}(r)\rangle_{RS}=\varepsilon_{\lambda}^{(0)}|\phi_{\lambda}(r)\rangle_{RS},\hspace{0.3cm}(\lambda=1,2,...)
\label{eq:basis}
\end{eqnarray}
and derive all the eigen-states $\{|\phi_{\lambda}(r)\rangle_{RS}|\lambda=0,1,2,...\}$ of $\hat{H}_{\rm eff}^\prime(E_g^{(0)})$. Notice that in Eq. (\ref{eq:basis})
the parameter $(E_g^{(0)}$ is already determined, rather than an unknown.
Then we 
diagonalize $\hat{H}_{\rm eff}(E_g^{(0)})$ in the basis $\{|\phi_{\lambda}(r)\rangle_{RS}|\lambda=0,1,2,...\}$. In our calculations we take into account all the eigen-states $|\phi_{\lambda}(r)\rangle_{RS}$ with $\varepsilon_{\lambda}^{(0)}<E_{\rm cut}$, where $E_{\rm cut}$ is about $8\omega$ and $\omega$ is the confinement frequency given in Eq.~(\ref{conff}). As a result, about $200$ of the eigen states of the c.m. Hamiltonian ${\hat H}_R$ defined in Eq.~(\ref{hr}) are included in our calculation.

Our above treatment is based on the following reasons. {\bf (i):} It is clear that all the eigen-states of $\hat{H}_{\rm eff}^\prime(E_g^{(0)})$ satisfy the BPC (\ref{BPC}). {\bf (ii):} Since in $\hat{H}_{\rm eff}^\prime$ the c.m. and relative motion are not coupled with each other, Eq.~(\ref{eq:basis}) can be solved easily. {\bf (iii):} Since the atoms are moving near the trap center, the difference between $\hat{H}_{\rm eff}(E_g^{(0)})$ and $\hat{H}_{\rm eff}^\prime(E_g^{(0)})$, i.e., the term $V_{1}^{\rm (c)}({\hat {\bf R}}, {\bf r})$ defined in Eq. (\ref{Hpert}), is a perturbation. Thus, the numerical diagonalization of $\hat{H}_{\rm eff}(E_g^{(0)})$ in the basis $\{|\phi_{\lambda}(r)\rangle_{RS}|\lambda=0,1,2,...\}$ can be performed efficiently.

We also use this method in the diagonalization of $\hat{H}_{\rm eff}(E_g^{(j)})$ ($j=1,2,...$), which are performed in the iterative calculation of Sec.~\ref{sec:sec3C}. Namely, for each $j$ we first derive the eigen-states of $\hat{H}_{\rm eff}^\prime(E_g^{(j)})$, and then diagonalize $\hat{H}_{\rm eff}(E_g^{(j)})$ in the basis formed by these states.

\section{Data Fitting and Uncertainty Estimation }
\label{sec:appD}

In this appendix we show our approach for the fitting 
of
our theoretical results to the experimental measurements of the bound-state energies as well as the method for the estimation of the uncertainty of the best-fitting values of $a_{\pm}$. Here we take the system of $^{173}$Yb atoms as an example to introduce our method.

We denote the experimental results of bound-state energies of $^{173}$Yb atoms in the lattice with $s_f=30$, which are given by Refs.~\cite{PRL2015,PRX2019}, as $\{(B^{(1)},E_b^{(1)}),(B^{(2)},E_b^{(2)}),\cdots,(B^{(N)},E_b^{(N)})\}$. Here $B^{(i)}$ and $E_b^{(i)}$ ($i=1,...,N$) are the value of the bias magnetic field $B$ and the measured bound-state energy of the $i$-th data. We further denote
the bound-state energy given by our theoretical calculation as $E_b^{\rm (theory)}(B,a_+,a_-)$, which is a function of the bias magnetic field and the scattering lengths $a_{\pm}$. Using the nonlinear least square method \cite{NLF}, we determine the best-fitting values of $a_{\pm}$, which are denoted as 
$a_{\pm}^{\rm(best\ fitting)}$,
 by 
minimizing the target function 
\begin{equation}
S=\sum_{i=1}^N \left[ E_b^{(i)} -E_b^{\rm (theory)}(B^{(i)},a_+,a_-) \right]^2.
\end{equation}
Furthermore, we estimate the uncertainty ($95\%$ confidence) of the best-fitting values of $a_\pm$, which are denoted as $\delta_\pm=1.96\sqrt{\sigma_r^2 \lambda_\pm}$ with $\sigma_r$ being defined as
\begin{equation}
\sigma_r^2=\frac{1}{N-2}\sum_{i=1}^N \left[ E_b^{(i)}-E_b^{\rm (theory)}\left(B^{(i)},a_+^{\rm (best\ fitting)},a_-^{\rm (best\ fitting)}\right)\right]^2,
\end{equation} 
and $\lambda_\pm$ being defined as the eigenvalues of $\mathbf{H}^{-1}$	 with $\mathbf{H}$ given by
\begin{eqnarray}
&&\mathbf{H}=\nonumber\\
&&\left. \left(\begin{array}{ccc}
\frac{\partial E_b^{\rm (theory)}\left(B^{(1)};a_{+},a_{-}\right)}{\partial a_{+}} & \cdots & \frac{\partial E_b^{\rm (theory)}\left(B^{(N)};a_{+},a_{-}\right)}{\partial a_{+}}\\
\frac{\partial E_b^{\rm (theory)}\left(B^{(1)};a_{+},a_{-}\right)}{\partial a_{-}} & \cdots & \frac{\partial E_b^{\rm (theory)}\left(B^{(N)};a_{+},a_{-}\right)}{\partial a_{-}}
\end{array}\right)^{\intercal}\cdot\left(\begin{array}{ccc}
\frac{\partial E_b^{\rm (theory)}\left(B^{(1)};a_{+},a_{-}\right)}{\partial a_{+}} & \cdots & \frac{\partial E_b^{\rm (theory)}\left(B^{(N)};a_{+},a_{-}\right)}{\partial a_{+}}\\
\frac{\partial E_b^{\rm (theory)}\left(B^{(1)};a_{+},a_{-}\right)}{\partial a_{-}} & \cdots & \frac{\partial E_b^{\rm (theory)}\left(B^{(N)};a_{+},a_{-}\right)}{\partial a_{-}}
\end{array}\right) \right|_{a_\pm=a_\pm^{\rm (best\ fitting)}}.\nonumber\\
\end{eqnarray}
\end{widetext}


\begin{thebibliography}{52}
\bibitem [{foo({\natexlab{c}})}]{footnote1}%
  \BibitemOpen
  \bibinfo {title} {Here, the characteristic length of the van der waals
  potential $\beta_6$ is twice as large as the van der waals length $r_{\rm
  {vdw}}$, i.e. $\beta_6 = 2 r_{\rm{vdw}}$}\ \BibitemShut {NoStop}%
\bibitem [{foo({\natexlab{d}})}]{footnote4}%
  \BibitemOpen
  \bibinfo {title} {\uppercase{T}he divergences of $a_{i,j}$ ($i,j=o,c$) in (c)
  may be understood as a generalization of the divergence of the scattering
  length $a(E)$ in the single-channel $s$-wave scattering problems, which
  occurs at the energies with corresponding $s$-wave phase shift
  $\delta_0=n\pi$ with $n$ being any integer}\ \BibitemShut {NoStop}%
\end{thebibliography}
\end{document}